# BigBrain-MR: a new digital phantom with anatomically-realistic magnetic resonance properties at 100-µm resolution for magnetic resonance methods development

Cristina Sainz Martinez,[1,2] Meritxell Bach Cuadra,[2,3] João Jorge [1,*]

[1] CSEM - Swiss Center for Electronics and Microtechnology, Switzerland;

[2] CIBM Center for Biomedical Imaging, Switzerland;

[3] Department of Radiology, Lausanne University Hospital and University of Lausanne, Switzerland

**Abstract**

The benefits, opportunities and growing availability of ultra-high field magnetic resonance imaging (MRI) for humans have prompted an expansion in research and development efforts towards increasingly more advanced high-resolution imaging techniques. To maximize their effectiveness, these efforts need to be supported by powerful computational simulation platforms that can adequately reproduce the biophysical characteristics of MRI, with high spatial resolution. In this work, we have sought to address this need by developing a novel digital phantom with realistic anatomical detail up to 100-µm resolution, including multiple MRI properties that affect image generation. This phantom, termed BigBrain-MR, was generated from the publicly available BigBrain histological dataset and lower-resolution in-vivo 7T-MRI data, using a newly-developed image processing framework that allows mapping the general properties of the latter into the fine anatomical scale of the former. Overall, the mapping framework was found to be effective and robust, yielding a diverse range of realistic "in-vivo-like" MRI contrasts and maps at 100-µm resolution. BigBrain-MR was then tested in three different imaging applications (motion effects and interpolation, super-resolution imaging, and parallel imaging reconstruction) to investigate its properties, value and validity as a simulation platform. The results consistently showed that BigBrain-MR can closely approximate the behavior of real in-vivo data, more realistically and with more extensive features than a more classic option such as the Shepp-Logan phantom. This novel phantom is therefore deemed a favorable choice to support methodological development in brain MRI, and has been made freely available to the community.

***Corresponding author:**

João Jorge, PhD

CSEM
Freiburgstrasse 2
3008 Bern
Switzerland

e-mail: joao.jorge@csem.ch

# 1. Introduction

The growing availability of ultra-high field (UHF) magnetic resonance imaging (MRI) systems is expected to bring substantial benefit to diverse applications in neuroscience research and clinical practice (Kraff et al., 2015). Among its most promising contributions, UHF-MRI can offer increased signal-to-noise ratio (SNR), as well as specific enhancements in sensitivity for certain contrast mechanisms, such as those based on magnetic susceptibility (Duyn, 2012), which underly some of the most important structural and functional modalities available for brain imaging (Haacke et al., 2004; Ogawa et al., 1993). The increased sensitivity can then be flexibly traded for shorter acquisition times, as well as for higher spatial resolution, greatly increasing specificity down to the level of cortical columns and layers, for example (Deistung et al., 2013; Dumoulin et al., 2018; Polimeni et al., 2010).

The promising benefits of UHF-MRI and its growing availability have prompted an expansion in research and development of increasingly more advanced UHF imaging techniques, and in improving their robustness to enable their translation to routine clinical practice (Trattnig et al., 2016). To be effective, these efforts can critically benefit from numerous theoretical and computational tools. Amidst the latter, a clear need exists for powerful computational simulation platforms that can adequately reproduce the biophysical characteristics and mechanisms underlying brain MRI, with high spatial resolution. While well-established tools already exist to simulate the evolution of a magnetization signal based on known MR properties (e.g. $T_1$ and $T_2$ relaxation, $B_0$ and $B_1$ field inhomogeneities) (Kwan et al., 1996; Benoit-Cattin et al., 2005), it is also crucial for imaging simulations to account for how these properties are distributed across the brain in a realistic manner. More precisely, an effective simulation platform for high-resolution brain MRI presents two important requirements:

1. The need for realistic anatomical detail at sub-millimeter scale, while covering the whole brain. Ideally, the spatial specificity should be substantially finer than the resolution limits of the imaging methodology under investigation, so as to better model features such as partial volume effects, sub-voxel behavior and motion.
2. The need to incorporate, at this fine scale, the diverse biophysical properties that affect the outcome of MR image acquisitions. This includes relaxation properties such as $T_1$, $T_2$, and $T_2$*, susceptibility-induced effects, $B_0$ and $B_1$ inhomogeneities.

The requirements described above for a simulation platform, or in other words, a digital phantom, can be tackled with different approaches. A number of fine-scale in-vivo (Federau and Gallichan, 2016; Lüsebrink et al., 2021) and ex-vivo (Edlow et al., 2019; Kim et al., 2021) MRI datasets are publicly available, obtained with long-duration acquisitions at UHF, and resolutions up to 100 µm (Edlow et al., 2019). Other efforts have instead focused on in-vivo data at lower-resolution, but with more diverse modalities, allowing the definition of probability maps for multiple tissue types (e.g. gray and white matter, dura, skull, etc.), for more flexible simulations at moderate resolution with diverse contrasts (Collins et al., 1998; Aubert-Broche et al., 2006; Cocosco et al., 1997). Such approaches do offer, by nature, realistic anatomical information for simulations, with realistic MRI properties. However, they also have important limitations: (i) the available resolution remains limited by prohibitively increasing acquisition time requirements, caused not only by the additional k-space readout steps but also by the need for more averaging repetitions, to achieve a usable SNR (Lüsebrink et al., 2017); (ii) linked to (i), most available datasets, especially at higher resolutions, only provide a limited repertoire of MRI properties for the same anatomical sample, (iii) the acquisitions will be affected by artifacts (e.g. motion and breathing in-vivo (van Gelderen et al., 2007; Lüsebrink et al., 2017), microscopic air bubbles and altered MR properties ex-vivo (Shatil et al., 2016; Tovi and Ericsson, 1992)), which can be reduced by dedicated methods (Maclaren et al., 2012; Jorge et al., 2020), but not fully suppressed; (iv) the use of parallel imaging techniques (Deshmane et al., 2012) to reduce acquisition time may introduce biases on the image properties, which could affect the validity of certain simulations (especially those investigating parallel imaging methods).

In contrast with real brain datasets, some simulation phantoms have been designed based on purely mathematical models. A classic example is the Shepp-Logan phantom, composed of a simple set of ellipsoids of different intensities (Shepp and Logan, 1974). Some more complex models have also been created to better approximate the anatomy and geometry of the brain (Guerquin-Kern et al., 2012; Ngo et al., 2016), including models with well mathematically-defined formulations in both the image and Fourier domains, for improved simulation of k-space-based acquisitions (Guerquin-Kern et al., 2012). Mathematical models can incorporate diverse MRI properties as required (Gach et al., 2008), offer full control over the noise incidence and properties, and can be continuously improved and built upon. However, by design, the models can only approximate the intricate complexity of brain anatomy up to a fixed level of detail; hence, although such models are defined in the continuous domain and can be discretized with arbitrarily high resolution, the actual level of fine-scale detail remains limited.

More recently, a high-quality whole-brain histological dataset named BigBrain has been collected at 20 µm resolution and made publicly available in digital format (Amunts et al., 2013). This ex-vivo histological brain image has excellent quality and tissue contrast, and substantial ongoing efforts have been dedicated to the development of software tools to increase its value, including dedicated structure segmentations (Paquola et al., 2021), non-linear warping to standard space (Amunts et al., 2013) and cross-modal integration (Amunts et al., 2019). However, as a potential platform for MRI simulations, BigBrain lacks a direct correspondence to MRI properties – its contrast is based on a staining for cell bodies, and is imaged by optical methods, not MRI (Amunts et al., 2013).

In the present work, we have sought to develop a new platform for brain MRI simulations, which can also be viewed as a digital phantom, offering high-quality realistic anatomical detail down to 100 µm, and including multiple MRI properties that affect image generation – thereby fulfilling the two main requirements described before, and overcoming important limitations of the existing phantoms. The fine-scale anatomical detail is derived from BigBrain; a new image processing framework was developed for mapping lower-resolution real in-vivo data (including e.g. $T_1$ and magnetic susceptibility maps, complex coil sensitivities, etc.) into the fine-scale anatomical space of BigBrain, resulting in novel, 100-µm resolution "in-vivo-like" images and maps of these MRI properties. This level of spatial specificity is substantially higher than the current resolution achieved with 7T MRI acquisitions in-vivo, and is therefore expected to provide an appropriate base for UHF simulations. Having established the framework and generated a series of property maps for the phantom, which we have named BigBrain-MR, we then investigated its properties, value and validity as a simulation platform, in three different applications:

1. Simulation of motion effects and interpolation errors – studying the importance of the source image resolution when simulating motion and then generating displaced (and downsampled) images via interpolation.
2. Super-resolution imaging based on sub-voxel shifting along the slice encoding direction between multiple 2D acquisitions – a popular approach to increase resolution beyond the possibilities of individual acquisitions (Yue et al., 2016).

3. Parallel imaging reconstruction of 3D acquisitions, employing different undersampling schemes and acceleration levels – an almost ubiquitous component of modern imaging in diverse applications, ranging from fast acquisitions of moving samples (Zhang et al., 2015), to high-resolution scans that would otherwise require prohibitively long acquisition times (Federau and Gallichan, 2016).

The behavior of BigBrain-MR in these simulations was compared to that of real in-vivo data obtained under similar conditions, serving as a reference, and also to a more conventional Shepp-Logan-based computational phantom (Gach et al., 2008).

# 2. Methods

This study was approved by the local ethics committee (KEK Bern) and involved the participation of 2 healthy adult volunteers, who provided written informed consent. The data from one of the participants was used to develop BigBrain-MR (section 2.1), and that of the other participant was used for testing (section 2.2). All in-vivo data were acquired on a 7T Magnetom Terra scanner (Siemens Healthineers, Erlangen, Germany), equipped with a single-channel transmit, 32-channel receive head coil (Nova Medical, Wilmington MA, USA). All phantom development, processing and simulation steps were performed in Python, combined with tools from SigPy (Ong and Lustig, 2019), ANTs (Avants et al., 2008), FSL (Jenkinson et al., 2012) and ITK-SNAP (Yushkevich et al., 2006). The resulting phantom dataset has been made available in Zenodo, at https://zenodo.org/record/7432527, including all generated images and maps.

## 2.1. BigBrain-MR development

The framework for mapping lower-resolution in-vivo MRI contrasts and maps into the fine-scale anatomy of BigBrain involved several methodological steps (Fig. 1), which are described in detail in the sections below. Briefly, each in-vivo dataset (described in 2.1.1) is first registered to the BigBrain anatomy (2.1.2). Then, for each of 20 regions of interest (ROIs) previously defined for BigBrain (2.1.3), a moving average and moving standard deviation procedure are applied to obtain region-specific local offset and local variation maps, respectively, for that contrast (2.1.4). Finally, the in-vivo contrast is mapped into the fine-scale anatomy of BigBrain at 100 µm, based on a voxel-wise modulation by the offset and variation maps of the in-vivo and BigBrain contrasts (2.1.6); this procedure includes a partial volume model for voxels at the interface between ROIs, to create more realistic transitions (2.1.5).

[see Fig. 1 on page 23]

### 2.1.1. Source data

**BigBrain:** The 100-µm cell-density image (16-bit) and classification volume ("cls", 8-bit) in Montreal Neurological Institute (MNI) International Consortium for Brain Mapping (ICBM) 152 space were obtained from the 2015 data release of BigBrain (https://bigbrainproject.org/).

**ICBM:** $T_1$-, $T_2$- and PD-weighted brain images in MNI space, as well as accompanying tissue probability maps for gray matter (GM), white matter (WM) and cerebrospinal fluid (CSF), were obtained from the ICBM 152 atlas, version 2009c (http://nist.mni.mcgill.ca/atlases/).

**In-vivo:** The in-vivo MRI data were acquired from one of the participants at 0.6-mm isotropic resolution using: (i) a 3D MP2RAGE sequence (Marques et al., 2010) with 154×192×192-mm (whole-brain) field-of-view, TE/$TI_1$/$TI_2$/TR = 4.94/800/2700/6000 ms, $\alpha_1/\alpha_2$ = 7°/5°, 240 Hz/Px bandwidth, 3× GRAPPA acceleration in the first phase encoding (PE) direction and 6/8 partial Fourier in both PE directions, and acquisition time (TA) of approximately 10 minutes, and (ii) a flow-compensated 3D multi-echo gradient-recalled echo (ME-GRE) sequence with 173×230×144-mm (whole-brain) field-of-view, four echoes at $TE_1/TE_2/TE_3/TE_4$ = 4.97/13.17/21.37/29.57 ms, TR = 34 ms, α = 10°, 250 Hz/Px bandwidth, 2× GRAPPA acceleration and 7/8 partial Fourier in both PE directions, TA ≈ 8 min. The MP2RAGE acquisition was reconstructed online to yield a $T_1$-weighted ($T_1$w) image and a $T_1$ map. The ME-GRE raw data were reconstructed offline using Python tools developed in-house together with SigPy, to obtain the four $T_2$*-weighted ($T_2$*w) magnitude and phase images, an $R_2$* map and offset ("$TE_0$") image, a background field map, a quantitative susceptibility map (QSM), and 32 complex coil sensitivity maps. The sensitivity maps were estimated using ESPIRiT (Uecker et al., 2014). After this, as done in previous work (Jorge et al., 2020), the coil sensitivities were combined with a virtual "body coil-like" channel estimated via block coil compression (Bilgic et al., 2017, 2016), and the full GRE images (magnitude and phase) were thereby reconstructed using SENSE (Uecker et al., 2014) with wavelet-based regularization. The $R_2$* map and offset image were estimated with a weighted least-squares fit of mono-exponential $T_2$* relaxation across echoes. The background field map was obtained from the phase data using V-SHARP after phase unwrapping (Li et al., 2011). Finally, QSM was performed using the single-orientation STAR-QSM approach (Wei et al., 2015). A bias field map was also estimated from the first-echo $T_2$*w magnitude image using FSL-FAST (Zhang et al., 2001).

#### 2.1.2. Pre-processing and registration

The in-vivo source images were first brain-extracted using ITK-SNAP, and then registered to BigBrain (MNI) space using non-linear Symmetric Normalization ("SyN", comprising affine + deformable transformations) from ANTs (Fig. 1, top-left). Notably, the ICBM brain images, not BigBrain, were used as registration template; this approach was found more robust, most likely because the ICBM images offer closer contrasts to those of the in-vivo MRI data, as well as a more similar brain geometry (e.g. narrower fissure spaces than in BigBrain). The BigBrain and ICBM anatomies were confirmed to be well aligned overall, thus permitting this approach.

#### 2.1.3. Region labeling

A set of 20 different ROIs were labeled for both the BigBrain and ICBM datasets by combining different segmentation sources and techniques (Table I). For ICBM, the cortical GM and WM, CSF, and cerebellar GM and WM labels were generated from the tissue probability maps, combined with information from the CerebrA atlas for ICBM 2009c (Manera et al., 2020) to separate these regions. A total of 11 subcortical ROIs (e.g. thalamus, caudate, putamen) were obtained from the ICBM atlas created by (Xiao et al., 2019), and 3 from CerebrA (basal forebrain, brainstem, ventral diencephalon). The pineal gland was manually segmented using ITK-SNAP.

For BigBrain, a first estimate of the GM and WM labels was obtained from the "cls" volume; the GM/WM boundary was then refined with a locally-adaptive thresholding approach, to accommodate the large-scale drifts in GM and WM intensity across BigBrain (Supp. Fig. 1). The cerebellar GM and WM were directly obtained from cls. The same 11 sub-cortical structures obtained from (Xiao et al., 2019) for ICBM were likewise available from the same work for the BigBrain anatomy. The superior part of the brainstem, the ventral diencephalon and pineal gland were obtained from the corresponding ROIs of the ICBM anatomy, aligned in the same space; the basal forebrain and inferior part of the brainstem were obtained by direct manual segmentation in ITK-SNAP. Finally, the CSF ROI was created to fill-in the ventricles and form a realistic envelope around the brain – optimized by visual inspection since no CSF is present in the BigBrain dataset. This was achieved using the ICBM CSF mask as starting point, adding non-labeled inner regions, excluding already labeled regions, and then applying morphological dilation and erosion steps to create a smooth envelope around the brain.

In both ICBM and BigBrain, most ROI labels were further tuned with morphological operations (binary erosion, dilation, opening, closing), including intensity-selective dilations and erosions where allowed by the contrast. The choice of operations and parameters for each ROI was set empirically so as to obtain optimal anatomical delineations.

[see Table I on page 30]

### 2.1.4. Offset and variation mapping

Having warped a given in-vivo image to BigBrain space, a procedure was then applied to obtain maps of ROI-specific local mean intensity (termed *offset*) and standard deviation (termed *variation*) for that in-vivo contrast (Fig. 1 top-right, Supp. Fig. 2). In more detail, for each voxel of each BigBrain ROI, the offset was computed as the average in-vivo image intensity in a selection of nearby voxels. The selected voxels had to obey three criteria: (i) being within a radius $W$ of the voxel of interest, (ii) belonging to the same ROI, and (iii) having their intensity within a certain defined range (in cases where the in-vivo contrast allowed clear differentiations between ROIs based on histogram inspection). Condition (i) defined the "local" nature of the maps, with $W$ controlling the level of spatial detail that is retained from the in-vivo data; a radius $W$ of 5 mm was chosen so as to afford some robustness against noise in the input, knowing that the in-vivo 7T data used for this work had sub-millimeter voxels. For condition (ii), the voxels were required to belong to the same ROI as the center voxel in both the BigBrain labelling and the ICBM labelling; this conservative approach was intended to better account for imperfections in the spatial alignment between the in-vivo anatomy and BigBrain (also considering local differences between the latter and the ICBM brain), and thereby reduce the inclusion of voxels from other brain regions, which would bias the estimates. Condition (iii) was also implemented to further mitigate such "leakage" effects.

The estimation was performed for every voxel in each of the 20 ROIs defined for BigBrain (Table I), and was also repeated to estimate the variation map, simply by replacing the mean estimate by a standard deviation (Supp. Fig. 2). Besides the input (in-vivo) images, the procedure was also applied to the BigBrain cell density contrast itself, to generate its offset and variation maps.

### 2.1.5. Partial volume model

Alongside the ROIs, the BigBrain-MR framework included a partial volume map at 100 µm to create smooth intensity transitions between those regions in the generated images (Supp. Fig. 3). To estimate this map, every border between any two neighboring ROIs was identified, and a "transition area" was delineated by expanding the border by a length $L$ into each of the neighboring ROIs; $L$ was chosen empirically: 200 µm for borders with the CSF, 400 µm for all others. Considering any two neighbor ROIs A and B, and a voxel $v$ from ROI A inside the transition area, the intensity of $v$ was considered to be a composition of contributions from both ROIs:

$$I^v = \rho_A^v \, I_A^v + (1 - \rho_A^v) \, I_B^v \qquad \text{Eq. 1}$$

where $I_A^v$ and $I_B^v$ are the (unknown) tissue intensities stemming from tissue A and B, respectively, and $\rho_A^v$ is the proportion of tissue from A in $v$, which is in the range $[0,1]$ and constitutes the parameter of interest for partial volume mapping. The necessary voxel intensities were taken from the BigBrain cell density image at 100 µm. To estimate $\rho_A^v$ , the intensity $I_A^v$ (and respectively $I_B^v$) was approximated as the intensity of the closest voxel to *v* that belongs to A (resp. B) but lies outside the transition area (and is therefore considered to be a "pure tissue" intensity). Since $I^v$ is known, Eq. 1 can then be solved for $\rho_A^v$. This approach, which we named *intensity-based model*, was found effective for all borders where the neighboring regions have sufficiently distinct intensity (e.g. putamen vs. WM, most of the cortical GM-WM interface). For region pairs without sufficient contrast, i.e. where $I_A^v \sim I_B^v$ (e.g. putamen vs. globus pallidus, some parts of the cortical GM-WM interface), however, the intensity model (Eq. 1) becomes less determined, and the resulting $\rho_A^v$ estimates were noisy. This would then result in noisy-looking, inaccurate borders in the final generated images, especially if the corresponding in-vivo input data *did* have a clear contrast between the regions in question. To avoid this issue, the intensity-based model was complemented with a *distance-based model*, where $\rho_A^v$ for a voxel *v* is given by:

$$\rho_A^v = \frac{d_B^v}{d_A^v + d_B^v} \qquad \text{Eq. 2}$$

where $d_A^v$ (and respectively $d_B^v$) is the distance to the closest voxel to *v* that belongs to A (resp. B) but lies outside the transition area. This model therefore does not depend on intensities and produces robust results regardless of the ROI contrast; conversely, because it is mainly driven by how the transition area is defined, the model is bound to be inaccurate when contrast does exist and the border can be depicted. We therefore sought to combine and leverage both methods by weighting their contribution for each voxel based on the existence/absence of contrast, i.e., on the absolute difference $|I_A^v - I_B^v|$, as follows:

$$\rho_A^v\,|_{combined} = \beta_A^v\,\rho_A^v\,|_{intensity} + (1 - \beta_A^v)\,\rho_A^v\,|_{distance} \qquad \text{Eq. 3}$$

where the weighting $\beta_A^v$ is estimated by a sigmoid function of the intensity difference:

$$\beta_A^v = \frac{1}{1 + exp\big(-a(\delta_{AB}^v - b)\big)} \qquad \text{Eq. 4}$$

where $\delta_{AB}^v = |I_A^v - I_B^v|$, and $a$ and $b$ were fixed parameters (same for every voxel) that controlled the sigmoid shape. The latter parameters were set empirically by visual inspection so as to favor the intensity-based model as far as possible, while still suppressing "noisy" regions where the contrast was poor (Supp. Fig. 3a).

### 2.1.6. Contrast mapping

Having obtained the offset and variation maps in BigBrain space for a given in-vivo input image (section 2.1.4), a contrast mapping procedure was then applied to generate an "in-vivo-like" image with realistic structural detail at 100-µm resolution (Fig. 1, bottom). The mapping procedure was performed for each voxel in each of the BigBrain ROIs (described in 2.1.3), and included a partial volume composition model for voxels in transition areas (2.1.5). For a voxel $v$ inside an ROI A and outside transition areas, the final image intensity $I_{iv}^v$ was computed as:

$$I_{iv}^v = (I_{cd}^v - \mu_{cd}^v)\frac{\sigma_{iv}^v}{\sigma_{cd}^v} + \mu_{iv}^v \qquad \text{Eq. 5}$$

where $I_{cd}^{v}$ is the intensity of the same voxel in the BigBrain cell density image (100 μm), $\mu_{cd}^{v}$ and $\sigma_{cd}^{v}$ are the offset and variation values from the cell density contrast, and $\mu_{iv}^{v}$ and $\sigma_{iv}^{v}$ are the offset and variation from the in-vivo contrast. Thus, Eq. 5 effectively generates the new contrast by removing the local offset from the BigBrain contrast, re-scaling to the in-vivo contrast variation, and adding the in-vivo offset. Given how the offset and variation maps are estimated (section 2.1.4), this approach thereby introduces all (ROI-specific) contrast features from the in-vivo data that are constant or vary over spatial scales larger than 5 mm (the window width $W$) as part of the offset $\mu_{iv}^{v}$, while retaining the finer features of the BigBrain image, modulated by the variation $\sigma_{iv}^{v}$.

For a voxel $v$ in ROI A and inside a transition area with respect to a neighbor ROI B, the image intensity $I_{iv}^{v}$ was computed as:

$$I_{iv}^{v} = \rho_{A}^{v}\, I_{iv}^{vA} + (1 - \rho_{A}^{v})\, I_{iv}^{vB} \qquad \text{Eq. 6}$$

where $\rho_{A}^{v}$ is the tissue proportion from ROI A given by the partial volume map (section 2.1.5), and $I_{iv}^{vA}$ (respectively $I_{iv}^{vB}$) is the image intensity for the closest voxel to $v$ that belongs to A (resp. B) but lies outside the transition area, with its intensity obtained through Eq. 5.

### 2.1.7. Exceptions

The contrast mapping procedure described above (and outlined in Fig. 1) was not applied to the complex coil sensitivity maps, background field and bias field maps, given the particular nature of these data. For these cases, the in-vivo input data was simply linearly registered to ICBM / BigBrain space (relying on magnitude images from the same acquisitions), resampled to 400-μm resolution (deemed sufficiently high, given their general smoothness), and adjusted to the anatomical boundaries of BigBrain (Supp. Fig. 4).

## 2.2. BigBrain-MR tests

### 2.2.1. Interpolation errors and source resolution

The goal of this test was to investigate the incidence of image intensity errors that can be introduced by interpolation when simulating motion of a (non-parametric) computational phantom, and how the error evolves depending on the resolution of the source and the output (Fig. 2, Test 1). To this end, the BigBrain 100-μm image (cell density contrast) was taken as source data, and lower-resolution versions were obtained by downsampling to 200, 400, 800, and 1600 μm (Fig. 2, Test 1, blue path). In a first test, each of these "source" versions was then submitted to a rotation of 45° in the axial plane, and regridded by trilinear interpolation. The resulting matrices were then further downsampled to 200, 400, 800, and 1600 μm (as far as allowed by the starting resolution), to yield the output data. In parallel, a reference version was obtained by rotating the 100-μm source image directly (no prior downsampling), then regridding by trilinear interpolation, and finally downsampling to 200, 400, 800, and 1600 μm (Fig. 2, Test 1, violet path). All downsampling steps were performed by voxel averaging. Each of the results from the simulated lower-resolution sources was then compared to the respective reference result from the 100-μm source (at the same end resolution) by computing their normalized root-mean-squared error (NRMSE), in percentage:

$$NRMSE = 100 \sqrt{\frac{\overline{(I_L - I_R)^2}}{var(I_R)}} \quad \text{Eq. 7}$$

where $I_L$ represents the image resulting from a lower-resolution source and $I_R$ the corresponding reference from the 100-µm source. The mean and variance operations were restricted to voxels from the brain (based on a previously generated mask). Besides the 45°-rotation in the axial plane, three other motion types were studied: 45°-rotation in the sagittal plane, 100µm-translation in the posterior-to-anterior direction, and 100µm-translation in the inferior-to-superior direction.

[see Fig. 2 on page 24]

### 2.2.2. Super-resolution imaging

Here, we investigated the behavior of BigBrain-MR when used to simulate a super-resolution (SR) approach for 2D multi-slice images that allows reducing the slice thickness based on multiple stacks acquired with systematic position shifts along the slice axis (Greenspan et al., 2002; Yue et al., 2016) (Fig. 2, Test 2). The performance of BigBrain-MR was compared to that of real in-vivo data; a more basic Shepp-Logan-based computational phantom was also added for comparison, to investigate whether BigBrain-MR does constitute a closer approximation to real data than more conventional phantoms. For this test, a resolution of 0.4×0.4×2.4 mm was considered for the acquisitions, to then obtain super-resolved estimates at 0.4×0.4×8.0 mm.

**In-vivo data:** 2D multi-slice GRE data (TR/TE = 1900/13ms) were acquired from a human participant at 7T, consisting of three 2D stacks of 33 axial slices at 0.4×0.4×2.4 mm resolution, each positioned with a 0.8 mm displacement in the slice direction. A separate stack at 0.4×0.4×0.8 mm resolution, 99 slices, was acquired to serve as ground truth.

**Phantom data:** The SR experiment performed in vivo was identically simulated in BigBrain-MR using one of its $T_2$*-weighted magnitude images (TE = 13ms), multiplied by the bias field, and downsampled from the original 100-µm resolution. The same simulation was also performed with a computational 3D Shepp-Logan phantom (Gach et al., 2008), as implemented by Tomoyuki Sadakane (https://github.com/tsadakane/sl3d), “Toft-Schabel” type. This phantom was normalized so as to better approximate the in-vivo data: its ellipsoid form was given physical dimensions close to those of the in-vivo brain overall (by visual adjustment), and the intensities of its ROIs were manually adjusted to more closely match the average $T_2$*-weighted intensities in GM, WM and CSF observed in vivo. To simulate the 2D acquisitions, a rectangular slice profile was adopted (Greenspan et al., 2002), which in this case amounted to simply averaging sets of 24 slices (respectively 8 slices for the ground truth) from the 100-µm source to obtain 2.4mm slices (respectively 0.8mm) (Supp. Fig. 5a).

**SR estimation:** The estimation of an 8mm-thick image $x$ from a set of three 24mm-thick acquisitions $Y_k$ was performed by minimization of the following cost function:

$$E(x) = \sum_{k=1}^{3} ||Y_k - H_k x||_2^2 + \lambda\, Reg(x) \quad \text{Eq. 8}$$

where the operator $H_k$ applies an appropriate displacement in the slice direction (a multiple of 0.8mm) followed by averaging of slices in sets of 3, and $Reg$ is a regularization function of $x$, with a weighting factor $\lambda$. Two different options were tested for regularization: (i) total variation (TV), based on the $L_1$- norm of the image gradient (Strong and Chan, 2003), and (ii) the $L_1$-norm of the Daubechies-4 wavelet transform (WT) (Lustig et al., 2007). SigPy (Ong and Lustig, 2019) was used to implement and iteratively solve the minimization problem (primal-dual hybrid gradient method for the TV-regularized approach, 1st-order gradient method for the WT case). Afterwards, the estimation performance for each case was evaluated based on the NRMSE (Eq. 7) between each estimated image and the respective ground-truth (as in 2.2.1, the mean was restricted to voxels inside the brain). In order to test all datasets (in-vivo and phantoms) with the same range of $\lambda$ values, each dataset was normalized by the mean value of $x$ prior to the SR estimation.

### 2.2.3. Parallel imaging reconstruction

In this test, we investigated the behavior of BigBrain-MR when used to simulate parallel imaging (PI) reconstructions, under different undersampling schemes and acceleration levels (Fig. 2, Test 3). A 3D imaging case was considered, with readout along the anterior-posterior direction and acceleration along one or both phase encoding directions. Four different undersampling schemes were tested: regular (Blaimer et al., 2004), CAIPIRINHA (hereafter named CAIPI) (Breuer et al., 2003), random (Lustig et al., 2007) and Poisson disk sampling (Vasanawala et al., 2011). For the regular and CAIPI cases, undersampling rates of 2×1, 2×2, 3×2 and 3×3 were tested; for the random and Poisson disk schemes, corresponding total rates of 2, 4, 6 and 9 were used. A calibration area of 16×16 lines at the center of k-space was kept fully sampled. For simplicity, no other acceleration methods (e.g. partial-Fourier undersampling) were considered. As in 2.2.2, the performance of BigBrain-MR was compared to that of real in-vivo data and a Shepp-Logan phantom.

**In-vivo data:** A fully-sampled 3D GRE sequence (TR/TE = 24/18ms) covering the whole brain at 1.4-mm isotropic resolution was acquired to serve as ground truth. As in 2.1.1, the complex sensitivity maps of its 32 receive channels were estimated directly from the data using ESPIRiT (Uecker et al., 2014), and the magnitude and phase image from the brain was estimated with the aid of block coil compression (Bilgic et al., 2017; Jorge et al., 2020). From the resulting fully-sampled complex data, the different undersampling cases were then simulated by retrospectively selecting the desired k-space readout lines (including the calibration area) and discarding the rest (Supp. Fig. 5b).

**Phantom data:** The PI experiment performed in vivo was identically applied to BigBrain-MR, downsampled to 1.4mm. For this test, a complex brain image was created using one of the $T_2$*-weighted magnitude images (TE = 21ms), multiplied by the bias field, and combined with an appropriate phase image, generated for the same TE based on the QSM and background field map of BigBrain-MR. The Shepp-Logan phantom was generated and normalized as for the SR test (section 2.2.2), with its phase component set to zero. Since Shepp-Logan does not provide coil sensitivity data, the complex sensitivity maps from BigBrain-MR were used for both phantoms. As for the in-vivo data, the different undersampling cases were simulated from the fully-sampled complex phantom images by retrospective selection of the corresponding k-space lines (Supp. Fig. 5b).

**PI reconstruction:** The estimation of the underlying non-aliased image $x$ from the set of 32 undersampled k-space acquisition channels $Y_k$ was performed with a wavelet-regularized SENSE approach (Lustig et al., 2007), by minimizing the following cost function:

$$E(x) = \sum_{k=1}^{32} ||Y_k - UFS_k x||_2^2 + \lambda \, Reg(x) \qquad \text{Eq. 9}$$

where the operator $S_k$ performs a multiplication by the respective sensitivity map, $F$ performs a Fourier transform, $U$ is the sampling mask, and $Reg$ implements the $L_1$-norm of the Daubechies-4 wavelet transform of $x$ (Lustig et al., 2007). SigPy (Ong and Lustig, 2019) was used to implement and iteratively solve the minimization problem (1st-order gradient method). After estimation, as in the SR test (section 2.2.2), the performance was evaluated based on the NRMSE (Eq. 7), with the fully-sampled images as ground-truth. In order to test all datasets with the same $\lambda$ range, each dataset was normalized by the mean value of $x$, and the coil sensitivity maps were normalized by $\sum ||S_k x||_2^2$, prior to reconstruction.

# 3. Results

## 3.1. BigBrain-MR development

In general, the processing framework developed for mapping lower-resolution in-vivo MRI data to the finer anatomy of BigBrain was found to perform effectively, resulting in images and maps with visibly realistic intensity and contrast properties, consistent with the source in-vivo data, yet with highly enhanced anatomical detail (Fig. 3a). The partial volume model included in the framework visibly enabled adequate transitions between brain regions, and between the brain and the background; it could be confirmed by visual inspection that both neighboring regions with different, as well as with similar intensities, were appropriately handled.

[see Fig. 3 on page 25]

While the mapping framework performed generally well, it did occasionally introduce small local imperfections in the generated images. These occurrences were most frequently caused by flaws in the region labeling, often in difficult regions of the cortex such as narrow sulci (Supp. Fig. 6a) and areas with unclear GM/WM boundaries (Supp. Fig. 6b). A few imperfections could also be attributed to the partial volume estimation approach (Supp. Fig. 6c). Apart from this, certain flaws present in the BigBrain histological preparation itself, such as e.g. local damage to the thalamus (Supp. Fig. 6d) and staining artifacts (Supp. Fig. 6e), were by design included in the generated images as well.

Altogether, the framework allowed the successful mapping of a diverse set of images and maps, including $T_1$- and $T_2$*-weighted images, and $T_1$, $R_2$* and susceptibility maps (Fig. 3b). Additional maps to simulate the background field, bias field and coil sensitivity maps (magnitude and phase) were also successfully generated for the same anatomical space (Fig. 3b), using the complementary approach described in 2.1.7. These maps contributed to considerably expand the simulation possibilities and realism offered by the phantom.

## 3.2. BigBrain-MR tests

### 3.2.1. Interpolation errors and source resolution

As expected, when simulating brain motion and generating the resulting images by interpolation, the starting resolution of the source image had a measurable impact on the precision of the result (Fig. 4). Overall, the rotations tested achieved larger NRMSE values than the translations. For the rotations, the error steadily increased with the voxel size of the source, while decreasing with the voxel size of the output. As a result, the highest errors were obtained at the largest source voxel size tested, 1.6 mm, with 1.6 mm output: 17.1% for the axial plane rotation, and 20.4% for the sagittal.

In the translation tests, for a given source resolution, the error also consistently decreased with the voxel size of the output. Unlike rotations, however, the error did not steadily increase with the source voxel size; for the range under analysis, the largest error was obtained at a source of 0.4mm: 6.4% for the posterior-to-anterior translation, and 8.7% for the inferior-to-superior – and then decreased again.

[see Fig. 4 on page 26]

### 3.2.2. Super-resolution imaging

Upon visual inspection, the simulations of lower-resolution slice-shifted acquisitions and higher-resolution ground-truth generated for BigBrain-MR and Shepp-Logan were in good agreement with the real measurements obtained in vivo (Supp. Fig. 5a). As expected for the chosen voxel dimensions, the lower-resolution acquisitions showed smoother borders between structures, especially in regions where these border surfaces were rapidly bending in the axial plane as a function of the slice height. This effect was well visible in numerous border regions of the gray matter and the ventricles for BigBrain-MR and in-vivo, and even more accentuated for the ellipsoid borders of Shepp-Logan (Supp. Fig. 5a).

For all three source datasets, and both regularization types, the estimation performance (quantified as the NRMSE with respect to the high-resolution ground truth) showed a clear dependence on the regularization weighting factor λ, with an initial decrease towards an optimal weighting, followed by a new increase past that optimal point (Fig. 5). Altogether, the implemented SR approach was found effective, achieving NRMSE values as low as 36% for the in-vivo dataset, 10% for BigBrain-MR and 4% for Shepp-Logan, whereas a more basic approach of simply up-sampling by linear interpolation could only achieve values of 47%, 24% and 32%, respectively.

In all three sources, the TV approach achieved a better performance (lower NRMSE) than WT (Fig. 5). Moreover, the behavior of NRMSE as a function of λ was more consistent across the three datasets with TV than with WT; in particular, the WT curve for the in-vivo data had a relatively similar shape to that of TV, whereas for the phantoms it varied substantially, showing a considerably smaller optimal weighting and a markedly more abrupt increase past that value. Besides this discrepancy with WT-regularization between the in-vivo case and the phantoms, another important difference observed in the tests was the scale of the NRMSE variations; for example, in the TV case, the NRMSE decrease from $\lambda = 0$ to the optimal value was found to be 41 to 36% for in-vivo, 18 to 10% for BigBrain-MR, and 13 to 4% for Shepp-Logan. Overall, between the two phantoms, BigBrain-MR consistently showed the closest behavior to the in-vivo case, including: (i) a closer range of NRMSE, (ii) a visibly more similar NRMSE behavior as a function of λ, with both TV and WT regularization, and (iii) substantially closer optimal regularization factors, which were found at 0.025 (TV) and 0.035 (WT) for the in-vivo case, 0.020 and 0.005 for BigBrain-MR, and 0.010 and 0.001 for Shepp-Logan.

[see Fig. 5 on page 27]

A direct visual inspection of the estimation results for different regularization weightings indicated that the NRMSE variations observed as a function of λ were consistent with conditions of under-, optimal and over-regularization in the image quality (Fig. 6). For all three sources, under-regularization led to results with a stronger degree of random noise, while over-regularization suppressed random noise but resulted in excessive smoothing of certain borders and features, along with a suppression of finer, but true, anatomical features. This important suppression effect occurred in vivo and was effectively captured by BigBrain-MR as well, but not by Shepp-Logan, which contained only a few basic shapes. The reconstructions performed with optimal weighting (i.e., at the NRMSE minimum) visually yielded the closest estimates to the respective ground truths in all three datasets, clearly sharper than the respective low-resolution acquisitions. This also confirmed the effectiveness of the implemented SR framework for recovering higher-resolution images.

[see Fig. 6 on page 27]

### 3.2.3. Parallel imaging reconstruction

The undersampled multi-channel measurements simulated for the in-vivo and BigBrain-MR data were found to produce considerably similar effects in both cases, including the aliasing patterns introduced by each undersampling scheme and the intensity modulations introduced by the spatially varying receive coil sensitivities. The same effects could also be observed in the Shepp-Logan phantom, albeit imposed on a simpler, considerably less anatomically realistic source sample (Supp. Fig. 5b).

For all three source datasets, and across all tested undersampling schemes and acceleration factors, the PI reconstruction performance (quantified as the NRMSE with respect to the ground truth) showed a clear dependence on the regularization weighting factor $\lambda$, with an initial decrease towards an optimal weighting, followed by a new increase past that optimal point (Fig. 7a). In all cases tested, the NRMSE consistently increased with the undersampling rate; at the same time, the dependence on regularization, i.e., the relative decrease in NRMSE achieved between $\lambda = 0$ and the optimal value, became visibly more accentuated. Additionally, the optimal weighting factor tended to decrease with the undersampling rate, except for the Poisson disk sampling scheme (Fig. 7b).

[see Fig. 7 on page 29]

While all three sources exhibited the above-described general patterns, deviations were present as well. The main difference observed in these tests resided in the scale of the NRMSE variations, which followed a similar trend to that observed in the SR tests; for instance, the minimum NRMSE across undersampling schemes and rates was at 8–34% for in-vivo, 4–22% for BigBrain-MR, and 5–11% for Shepp-Logan (Fig. 7a). Between the two phantoms, BigBrain-MR consistently showed the closest behavior to the in-vivo case, including: (i) a closer range of NRMSE, (ii) a more similar impact of regularization (see e.g. the 3×3 regular undersampling scheme curves for $\lambda$ between 0 and the optimum, Fig. 7a), and (iii) substantially closer optimal regularization factors (Fig. 7b).

As observed for SR, a direct visual inspection of reconstruction results for different regularization weightings indicated that the NRMSE variations observed as a function of $\lambda$ were consistent with conditions of under-, optimal and over-regularization in the image quality (Fig. 8). For all three sources, under-regularization led to results with a stronger degree of random noise, while over-regularization suppressed random noise but resulted in more “cartoonish” images, with excessive smoothing of certain anatomical features, and in some cases the accentuation of structured, yet artifactual patterns. The reconstructions performed with optimal weighting (i.e., at the NRMSE minimum) did visually yield the closest estimates to the respective ground truths.

[see Fig. 8 on page 29]

# 4. Discussion

In this work, we have developed a novel platform for brain MRI simulations, comprising a digital phantom with high-quality realistic anatomical detail up to 100-μm resolution, and including multiple MRI properties that affect image generation. This phantom, termed BigBrain-MR, was generated from the BigBrain histological dataset (Amunts et al., 2013) and a series of lower-resolution in-vivo 7T MRI data, using a newly-developed image processing framework that allows mapping the general properties of the latter into the fine anatomical scale of the former (Fig. 1). This framework was found to be effective overall, yielding a diverse range of "in-vivo-like" MRI contrasts and maps at 100 μm (Fig. 3). Following its development, BigBrain-MR was tested in three different imaging applications, to investigate its value and validity as a simulation platform (Fig. 2). Overall, the phantom was found to behave relatively similarly to real in-vivo data, and substantially more so than a more conventional Shepp-Logan phantom (Fig. 5 – Fig. 8). The advantages of working with a high-resolution source to simulate effects such as head motion were also well evinced (Fig. 4).

## 4.1. BigBrain-MR framework

Overall, the contrast mapping framework was found to work effectively and robustly, allowing the generation of diverse contrasts and maps that were found to be of very good quality overall, upon visual examination (Fig. 3). It is likely that this robustness is, to some extent, aided by the relative simplicity of the framework, which relies on three key modules: the mapping of offset and variation maps (2.1.4), the contrast mapping model (2.1.6), and the partial volume model (2.1.5). Altogether, this framework is mainly regulated by two parameters: (i) the radius of the spherical window used for local offset and variation mapping, $W$, described in 2.1.4, and (ii) the length $L$ defining the "transition area" between neighboring regions for the partial volume model, described in 2.1.5. As previously noted, $W$ controls the scale of anatomical information that is extracted from the input image (Supp. Fig. 2); a smaller radius will preserve finer-scale changes in the image properties (mean intensity and variability) within each region, but at the same time will be more sensitive to spurious contributions as well, such as from image noise and from "cross-region leakage" effects due to imperfect registration. For the input data explored in this work, the chosen radius of 5 mm was found to yield an adequate trade-off, and result in quality maps for all tested images. Naturally, despite the observed robustness, it is entirely possible that new input data may benefit from dedicated adjustments in this parameter, especially if the source resolution is markedly different from that explored in this work. Additionally, different values could also be employed for the offset and variation estimates, increasing the flexibility of the approach. Regarding the partial volume model, the parameter $L$ controls the spatial extent of what we termed "transition area", whose voxels are assumed to potentially contain tissue contributions from both interfacing regions (Supp. Fig. 3); a larger value (wider transition area) will in general ensure a more thorough coverage of the interfaces, as well as a more conservative selection of the reference "pure tissue" intensities (i.e., $I_A^v$ and $I_B^v$ in Eq. 1), which will thereafter be picked from voxels deeper within the ROIs (beyond the transition area); the downside to a larger $L$ is a heavier computational cost in the model estimation, and potentially more imperfections in the estimation for "tight" interface areas such as narrow cortical sulci (e.g. Supp. Fig. 6a), where there are multiple borders in close proximity. The values chosen in this work (200 μm for borders where one of the ROIs is CSF, 400 μm for all others) were found appropriate for the tissue properties and geometry of BigBrain, barring occasional local imperfections (Supp. Fig. 6c). The smaller $L$ value for borders with CSF was found sufficient for the generally sharp transitions between tissue and background in this dataset, while more effectively avoiding problems at the cortical sulci. As with the parameter *W*, the framework remains flexible with respect to *L*, and different values could be explored if new source datasets different from BigBrain are introduced.

Despite its general effectiveness, the proposed mapping framework does present noteworthy limitations. For instance, across different ROIs, the approach does effectively capture the differences that are expected for different image contrasts – e.g., in the $R_2$*-like map both the putamen and globus pallidus are hyperintense with respect to WM, while in the $T_1$w-like image only the putamen differs from WM (Fig. 3b). Within each ROI, however, the finer-scale features/textures are bound to be similar for all generated contrasts, by design, since they are all derived from the BigBrain histological image. For this reason, clearly, the proposed phantom is not intended to support any studies of the fine-scale relationship between image contrast and anatomical (micro-)structure – it is effectively meant to be a tool for imaging methods development. Also importantly, because BigBrain is a scalar image, the framework, in its current form, cannot generate maps of tensorial properties such as related to diffusion, flow or susceptibility. While the former limitation (contrast within ROIs) could potentially be overcome in a relatively straightforward manner by using a multi-contrast high-resolution source (e.g. histological data with multiple stainings from the same brain), the latter (tensor mapping) would likely require more dedicated extensions to the framework methodology. Another limitation of this design resides in the partial volume model, which was built to consider interfaces of only two ROIs for any given position. While this was indeed the case for the larger part of the (relatively coarse) set of 20 ROIs defined in this work, there were still select areas where the assumption was less valid (e.g. tight interface areas including the caudate, thalamus, WM and CSF, Supp. Fig. 3b); the difficulty in establishing a transition area in these and other geometrically-challenging regions also led to mapping imperfections in some cases (Supp. Fig. 6c). Importantly, we note that the proposed framework benefits from a relatively modular architecture (Fig. 1); its different steps can be independently improved, extended, or even replaced with relative ease, which may allow overcoming or mitigating the afore-mentioned limitations in future efforts, depending on the application needs.

Besides the mapping framework itself, certain flaws in the generated images could be attributed to the source data itself, including issues related to imperfect ROI labeling (Supp. Fig. 6a,b) and to artifacts from the histological preparation (Supp. Fig. 6d,e). Naturally, future work could be dedicated to improving the labeling, and going to even higher spatial resolutions could help in this regard as well (e.g. for finer sulci delineation). Regarding the histological imperfections, their incidence is clearly far from outweighing the remarkable value of this dataset, and the overall quality of the source and the generated images. One other, potentially more important limitation of the BigBrain dataset, however, is the lack of certain brain and non-brain structures that can play central roles in some methodological applications. Key examples include the brain vasculature, the skull, skin, fat layers, muscles and eyes – these structures can be of interest in themselves, or may be important sources of artifacts in certain imaging modalities, or play important roles in image processing tasks such as segmentation and registration, and therefore should be accounted for in the respective simulations. Importantly, while we adopted BigBrain as the source of fine-scale anatomical information for this work, the proposed mapping framework is flexible with respect to the source, and thus could be used with other datasets as well, depending on the application needs. Another interesting idea could be to attempt to insert and conciliate additional structures with the BigBrain space and anatomy, such as vasculature maps from other high-resolution sources (Bollmann et al., 2022; Duvernoy, 1999). This could potentially allow valuable future improvements and extensions to the phantom, while retaining its current advantages.

## 4.2. BigBrain-MR validation

Although far from covering all the possible applications of such a computational phantom, the three tests explored in this work (Fig. 2) provided diverse valuable insights with regard to the properties, practical value and validity of BigBrain-MR as a simulation platform. Test 1 was dedicated to the study of motion simulations, and the resulting errors introduced by interpolation. In in-vivo MRI, head motion during acquisitions cannot be fully avoided, and its impact becomes increasingly critical as higher resolution acquisitions are pursued (Jorge et al., 2018; Maclaren et al., 2012). For this reason, extensive research has been dedicated to the development of effective imaging methods to monitor and compensate for the effects of motion, and for which simulations can be extremely valuable (Zaitsev et al., 2015). The results obtained in Test 1 demonstrate, very clearly, that the resolution of the available source data and of the desired output images can have a strong impact on the introduction of errors with interpolation, which may prove large enough to affect the validity of the simulation results, depending on the application and methodology. Notably, the rotation case produced the strongest errors in the analysis (up to 20% of the intrinsic variations in the image); this may be due to the fact that the tested rotations introduced a continuous range of displacements across the image voxels (scaling with the distance of each voxel to the rotation axis) in two dimensions (the plane of rotation), whereas the tested translations only introduced a displacement along one dimension of the grid. It is important to note that the reference adopted for NRMSE estimation was not an absolute ground truth, but the “next best” available option – the 100 μm-resolution source data, which also suffered from interpolation errors in the rotation case (the translation, of 100 μm magnitude along one of the axes, was exempt from interpolation effects in practice). Despite this caveat, the very regular increase in NRMSE with the source resolution for the rotations indicates that the above choice was indeed the most accurate, and that valid interpretations can be drawn. Overall, the results strongly confirm that this type of simulations will benefit from using source data at the highest possible resolution, and the introduced errors will also be mitigated by setting for a relatively lower output resolution, whenever possible. In the future, extended tests could potentially explore whether such errors can be further minimized using higher-degree interpolation (e.g. cubic splines) and/or other downsampling approaches (e.g. k-space cropping after non-uniform Fourier transform) – though likely with significant added computational costs.

Tests 2 and 3 explored two extensively researched methodological applications: super-resolution imaging and parallel imaging reconstruction, respectively (Fig. 2). In both tests, all three source datasets behaved in generally good agreement with the respective theoretical expectations, on both the acquisition (Supp. Fig. 5) and estimation side (Fig. 6, Fig. 8). Nonetheless, in both tests, the behaviors observed in real in-vivo data were visibly better approximated by BigBrain-MR than by the popular Shepp-Logan phantom, which consistently tended to show less realistic behavior. For the dependence of the estimation performance on the regularization term (Fig. 5, Fig. 7), this outcome can likely be related to the intrinsic spatial properties of the two phantoms: Shepp-Logan is by design a piece-wise smooth object, which is exactly the type of property that is favored by the TV and WT-based $L_1$-norm regularization terms under investigation; BigBrain-MR is visibly more complex, and more akin to a real brain. Still, given that the optimal regularization weighting factors found in vivo were not perfectly matched by BigBrain-MR, the latter should not be expected to allow predicting the exact optimal value for a given application; nonetheless, its behavior with respect to the weighting and other parameters may likely prove extremely valuable to guide development efforts, test hypotheses and narrow down possible candidates for in-vivo testing. Beyond these advantages, BigBrain-MR also benefits from a realistic phase distribution, with both inner and outer brain contributions (background field), and from the availability of realistic coil sensitivity and bias field maps, which are necessary for diverse applications. Shepp-Logan does not include such features, and had to "adopt" the respective maps from BigBrain-MR for this work. Of note, despite focusing on 7T imaging, we do acknowledge that the voxel sizes used in Tests 2 and 3 were relatively large – this choice was necessary to enable acquiring in-vivo ground-truth data (super-resolved and fully k-space-sampled images, respectively) in manageable time. We do not, however, anticipate any reasons to expect different outcomes at higher resolutions. Overall, the tests suggest that BigBrain-MR constitutes a valuable and valid phantom for MRI methodological development, and a favorable choice with respect to existing phantoms such as Shepp-Logan.

# 5. Conclusion

This work proposes a novel computational brain phantom with high-quality realistic anatomical detail up to 100-μm resolution, named BigBrain-MR, which includes multiple MRI properties that affect image generation. The framework developed to generate this phantom was found effective and robust, and can be flexibly expanded in future work depending on application needs. A series of tests indicate that BigBrain-MR can closely approximate the behavior of real in-vivo data, more realistically and with more extensive features than the widely-used Shepp-Logan phantom.

# Acknowledgments

This work was funded by the Swiss National Science Foundation through grant 185909, and supported by CSEM – Swiss Center for Electronics and Microtechnology, by the Swiss Institute for Translational and Entrepreneurial Medicine (SITEM), and by the expertise of CIBM Center for Biomedical Imaging, Switzerland.

# Disclosure/conflicts of interest

The authors have no conflicts of interest to declare.

# Data availability statement

As part of the main goals of this work, the BigBrain-MR dataset generated by our mapping framework has been made publicly available in Zenodo, at https://zenodo.org/record/7432527.

# Figures

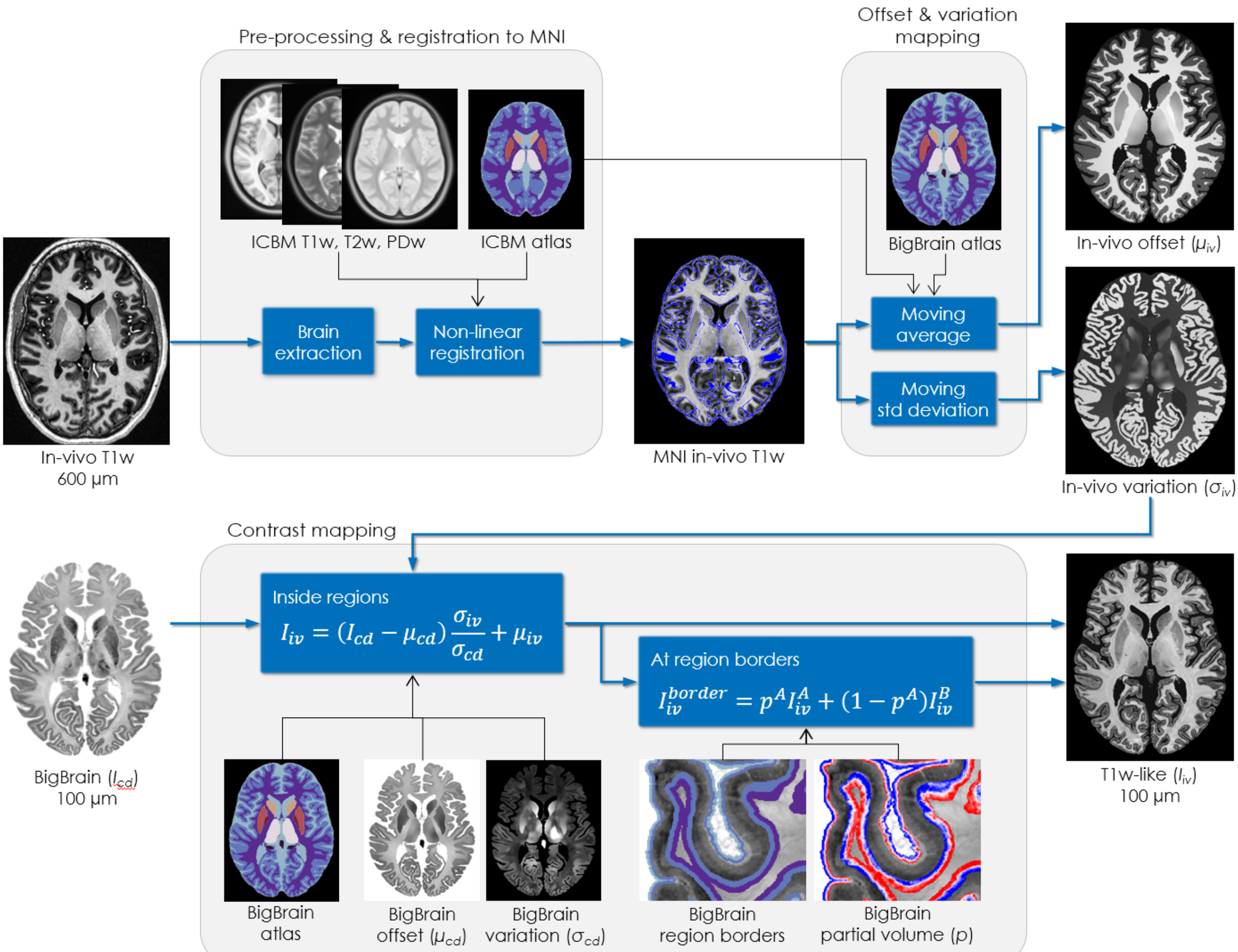

Fig. 1. Schematic outline of the BigBrain-MR framework for mapping lower-resolution in-vivo MRI contrasts and maps into the finer-scale anatomy of the BigBrain histological image, at 100-µm isotropic resolution. In this example, the in-vivo input is a $T_1$-weighted image acquired at 600-µm isotropic resolution.

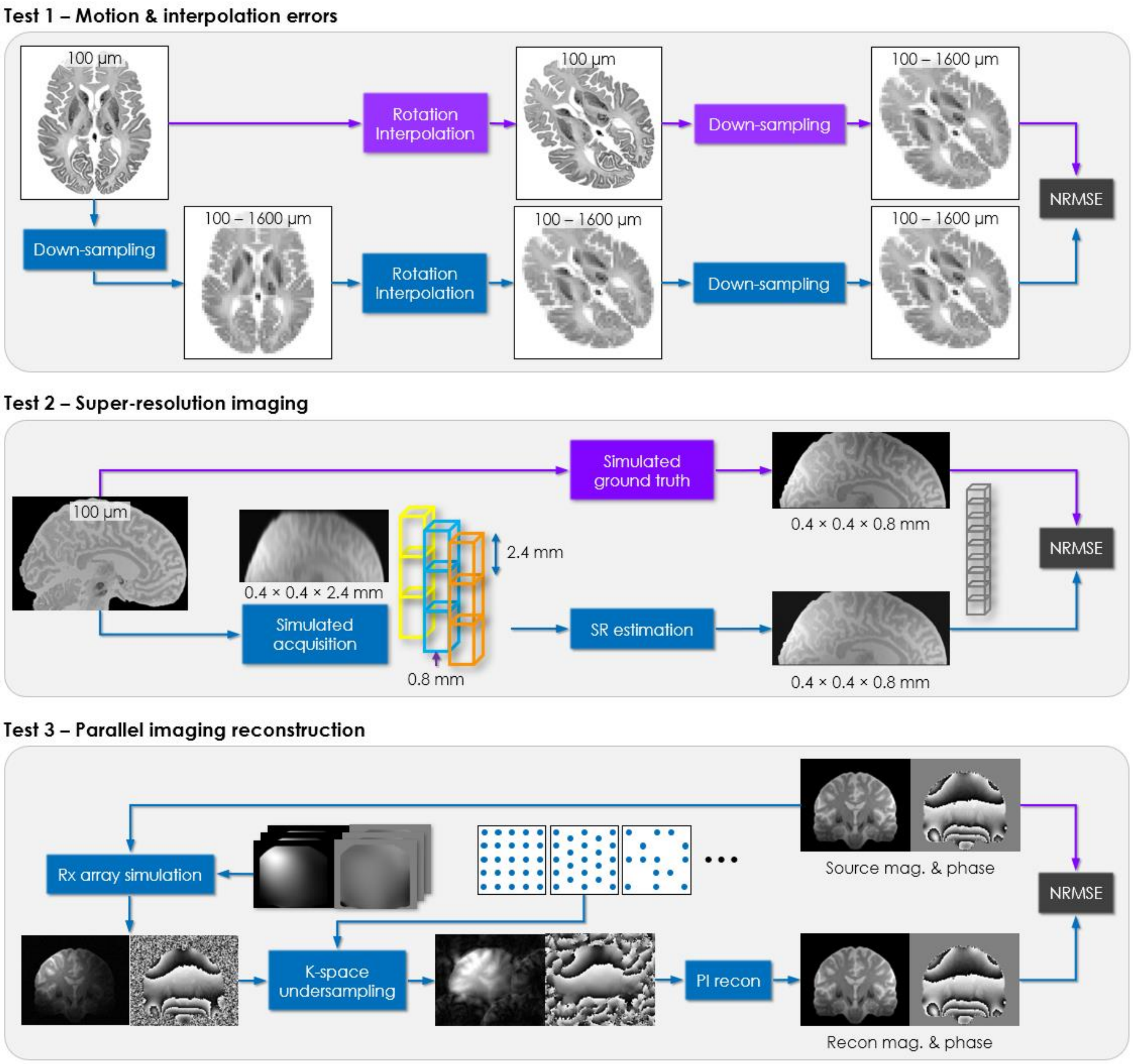

Fig. 2. Schematic outline of the three different applications explored in this work to investigate the properties, value and validity of BigBrain-MR as a simulation platform. For Test 2 and Test 3, the simulation procedures were also applied to a computational Shepp-Logan phantom. For the comparisons to in-vivo data, in Test 2, real acquisitions were performed to obtain both the shifted set of thicker-slice images and the thinner-slice ground-truth image, with all other parameters kept equal; for Test 3, an in-vivo fully-sampled 3D image was acquired, and then retrospectively undersampled in k-space.

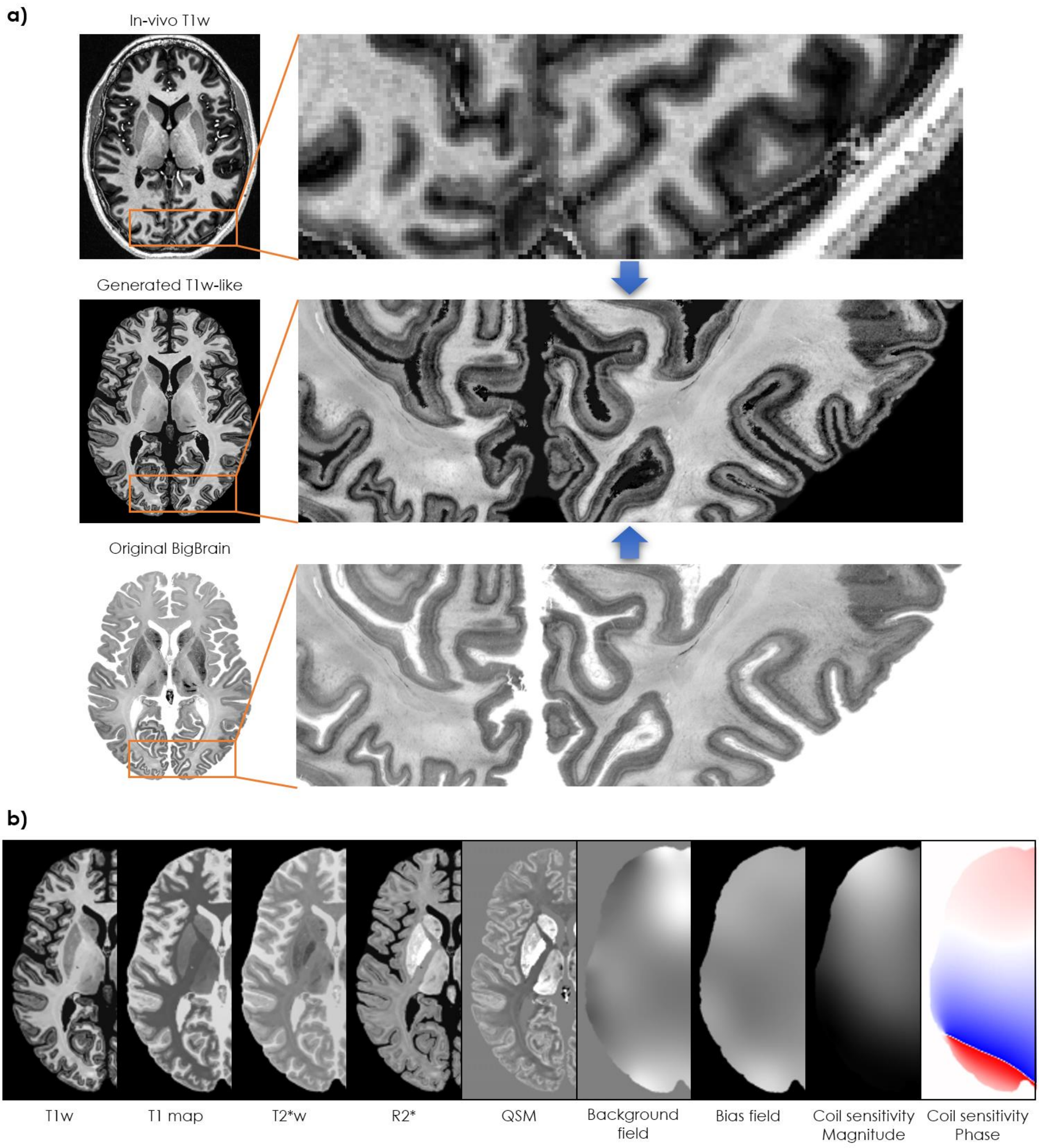

Fig. 3. MR-realistic dataset generated by the BigBrain-MR mapping framework at 100 µm resolution. a) Example of the fine-scale detail of an image generated from in-vivo $T_1$-weighted data originally acquired at 600 µm resolution. b) Illustration of the diverse contrasts and properties mapped in this work, including $T_1$- and $T_2$*-weighted images, $T_1$ and $R_2$* maps, magnetic susceptibility (QSM), background static magnetic field contribution, image bias field, and complex coil sensitivity maps (32-channel receive array).

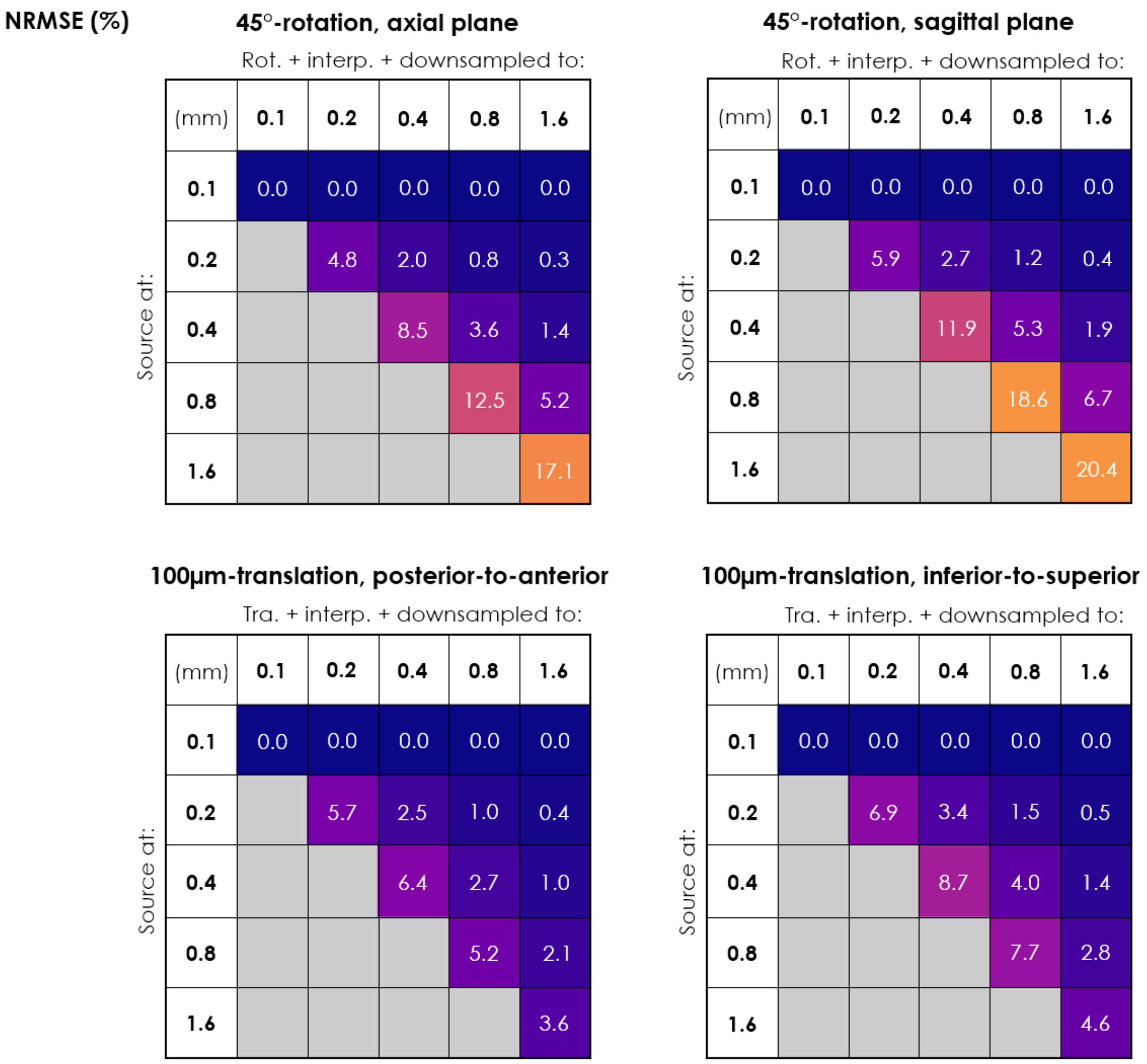

Fig. 4. Image intensity errors introduced by interpolation when simulating motion, and their dependence on the resolution of the source data and that of the output. The source and output resolutions are indicated in mm. The error is expressed as the normalized root-mean-squared error (NRMSE), in percentage. The error is computed with respect to the case that uses 100-µm source data (i.e. no downsampling before motion and interpolation, only after) – this case is itself also included in the matrices (1st row, thereby yielding null errors).

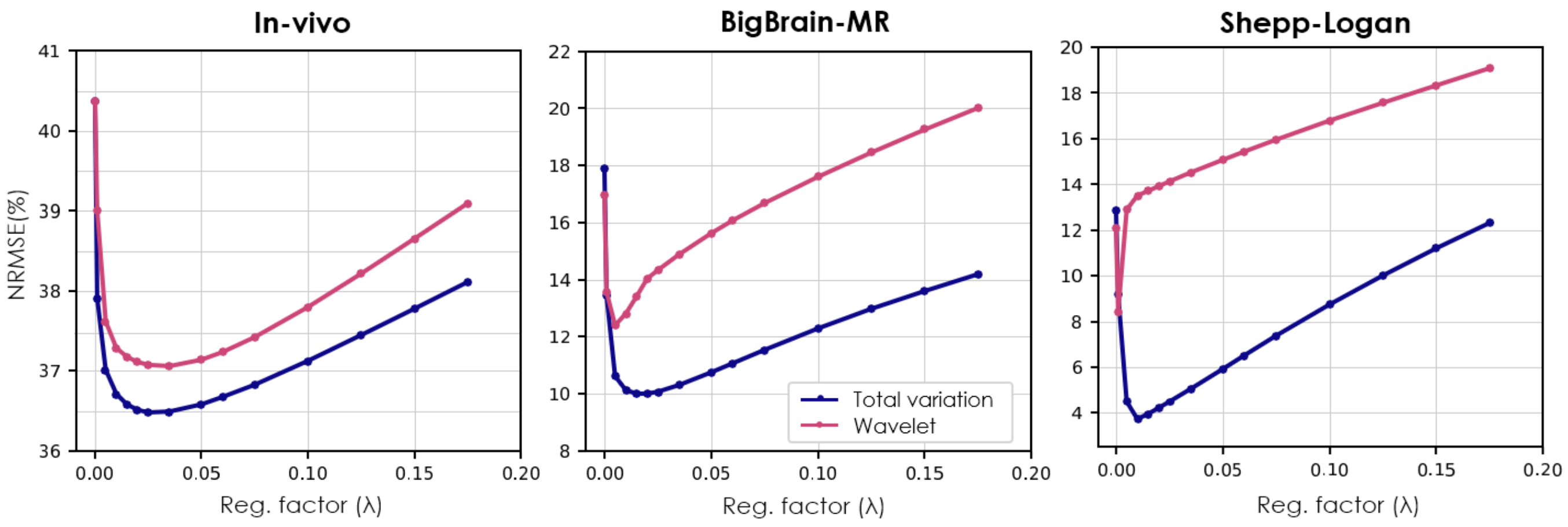

Fig. 5. Super-resolution estimation performance for different sources (in-vivo and digital phantoms) and different regulation approaches (total variation and wavelet decomposition), shown as a function of the weighting factor λ of the regularization term. The performance is quantified as the root-mean-squared error with respect to the ground truth image, normalized by the standard deviation of the ground truth, in percentage (NRMSE).

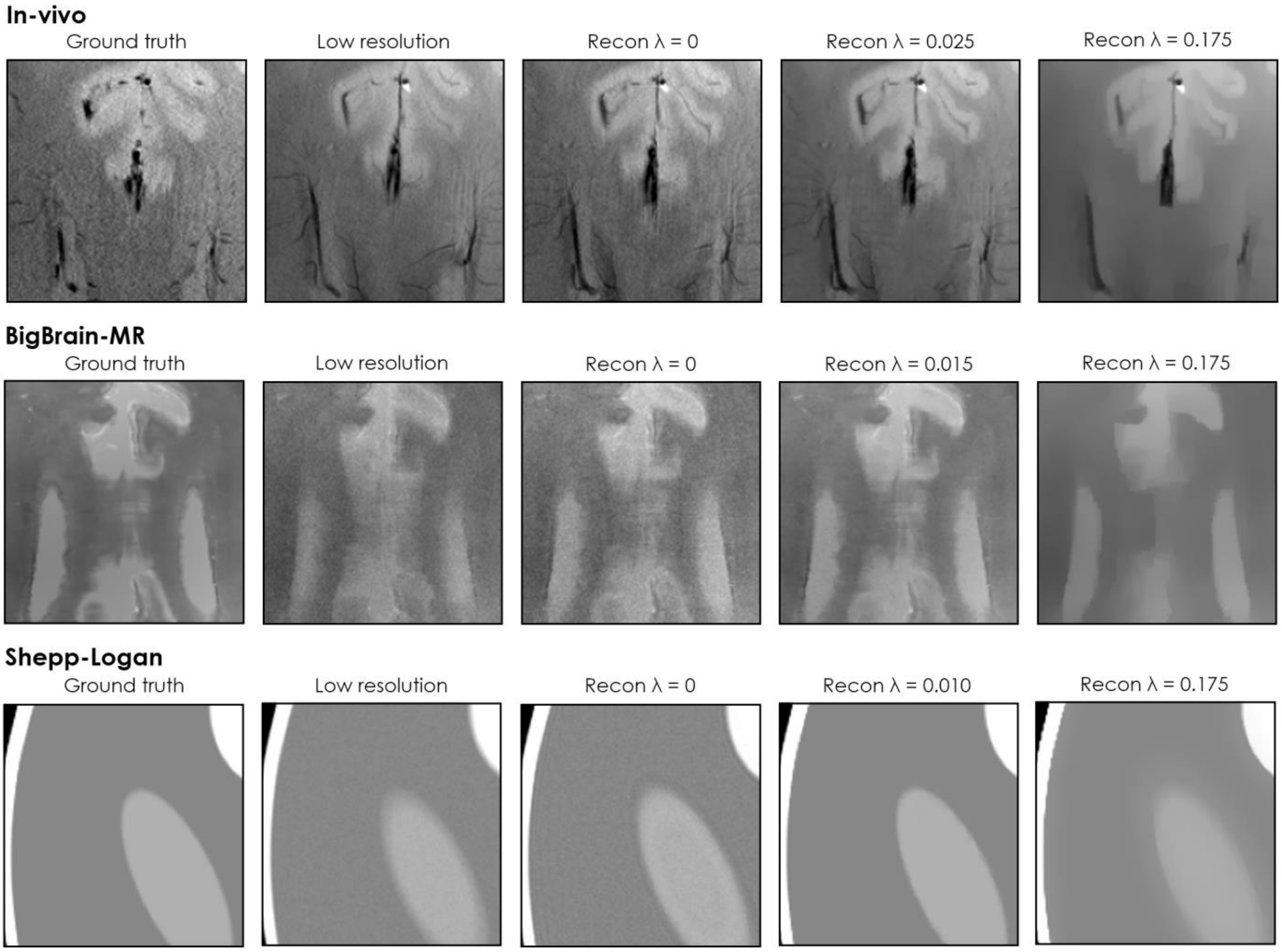

Fig. 6. Examples of super-resolution estimates for the in-vivo, BigBrain-MR and Shepp-Logan source datasets. The examples were chosen based on the NRMSE results of the SR simulations, and represent cases of under- (3rd column), optimal (4th column) and over-regularization (5th column). The high-resolution ground truth (1st column) and one of the low-resolution acquisitions used as input for SR (2nd column) are shown as well for comparison.

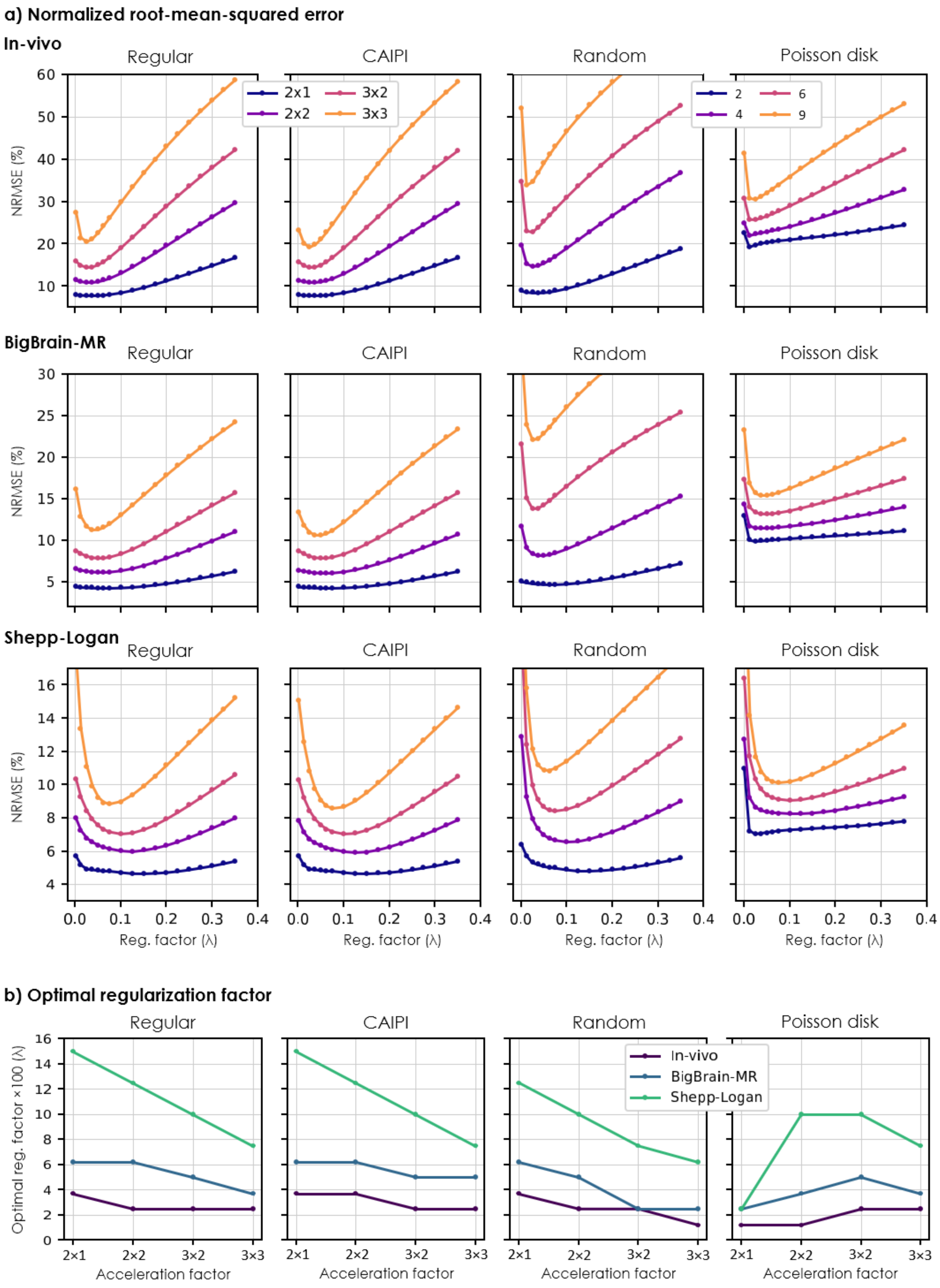

Fig. 7. a) Parallel imaging reconstruction performance for different source datasets (in-vivo and digital phantoms), undersampling schemes and acceleration factors, shown as a function of the weighting factor λ of the wavelet-based regularization term in the SENSE reconstruction. The performance is quantified as the root-mean-squared error with respect to the fully sampled image, normalized by the standard deviation of the fully sampled image, in percentage (NRMSE). b) Optimal weighting factor λ found for each of the tests reported in (a), corresponding to the minimum point of each of the NRMSE curves.

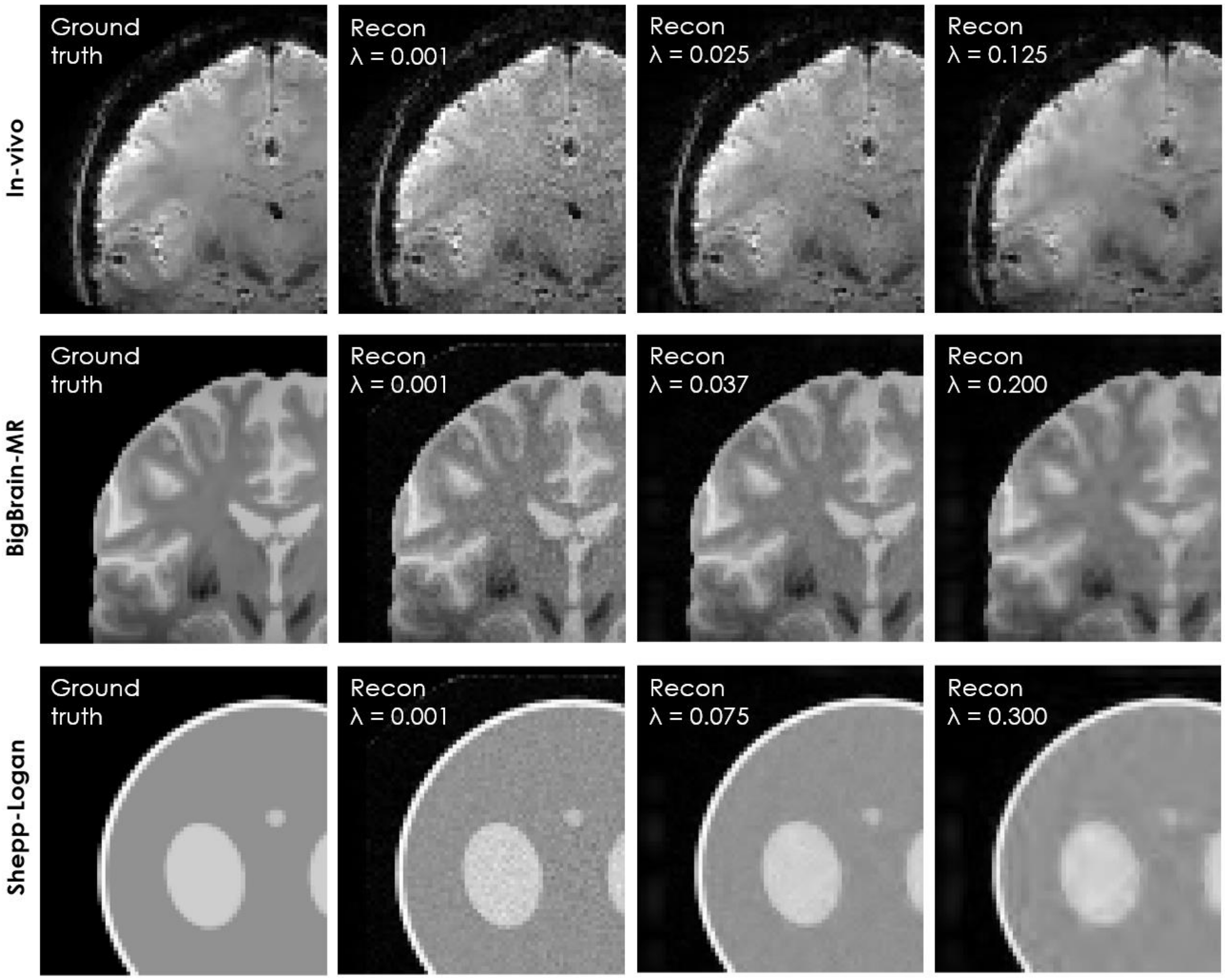

Fig. 8. Examples of parallel imaging reconstructions for the in-vivo, BigBrain-MR and Shepp-Logan datasets using 3×3 CAIPI undersampling and different weightings of wavelet-based regularization (λ). The examples were chosen based on the NRMSE results of the PI simulations, to represent cases of under- (2nd column), optimal (3rd column) and over-regularization (4th column). The fully-sampled ground truth is shown as well for comparison (1st column).

# Tables

Table I. Labeled regions of interest in ICBM and BigBrain anatomy

| Label | Name | ICBM source | BigBrain source |
|---|---|---|---|
| 1 | White matter | a | d + e |
| 2 | Gray matter | a | d + e |
| 3 | Cerebrospinal fluid | a + b | a + b |
| 4 | Cerebellum white matter | a + b | d |
| 5 | Cerebellum gray matter | a + b | d |
| 6 | Thalamus | c | c |
| 7 | Caudate | c | c |
| 8 | Putamen | c | c |
| 9 | External pallidum | c | c |
| 10 | Internal pallidum | c | c |
| 11 | Basal forebrain | b | g |
| 12 | Accumbens | c | c |
| 13 | Brainstem | b | f + g |
| 14 | Hippocampus | c | c |
| 15 | Ventral diencephalon | b | f |
| 16 | Amygdala | c | c |
| 17 | Subthalamic nuclei | c | c |
| 18 | Red nuclei | c | c |
| 19 | Substantia nigra | c | c |
| 20 | Pineal gland | g | f |

a. ICBM tissue probability maps (http://nist.mni.mcgill.ca/atlases/)
b. ICBM CerebrA labels for ICBM 2009c (Manera et al., 2020)
c. ICBM and BigBrain subcortical atlas (Xiao et al., 2019)
d. BigBrain classification volume (“cls”, https://bigbrainproject.org/)
e. Locally-adaptive GM/WM thresholding (Supp. Fig. 1)
f. Corresponding ICBM label
g. Manual segmentation (ITK-SNAP (Yushkevich et al., 2006))

# Supplementary Figures

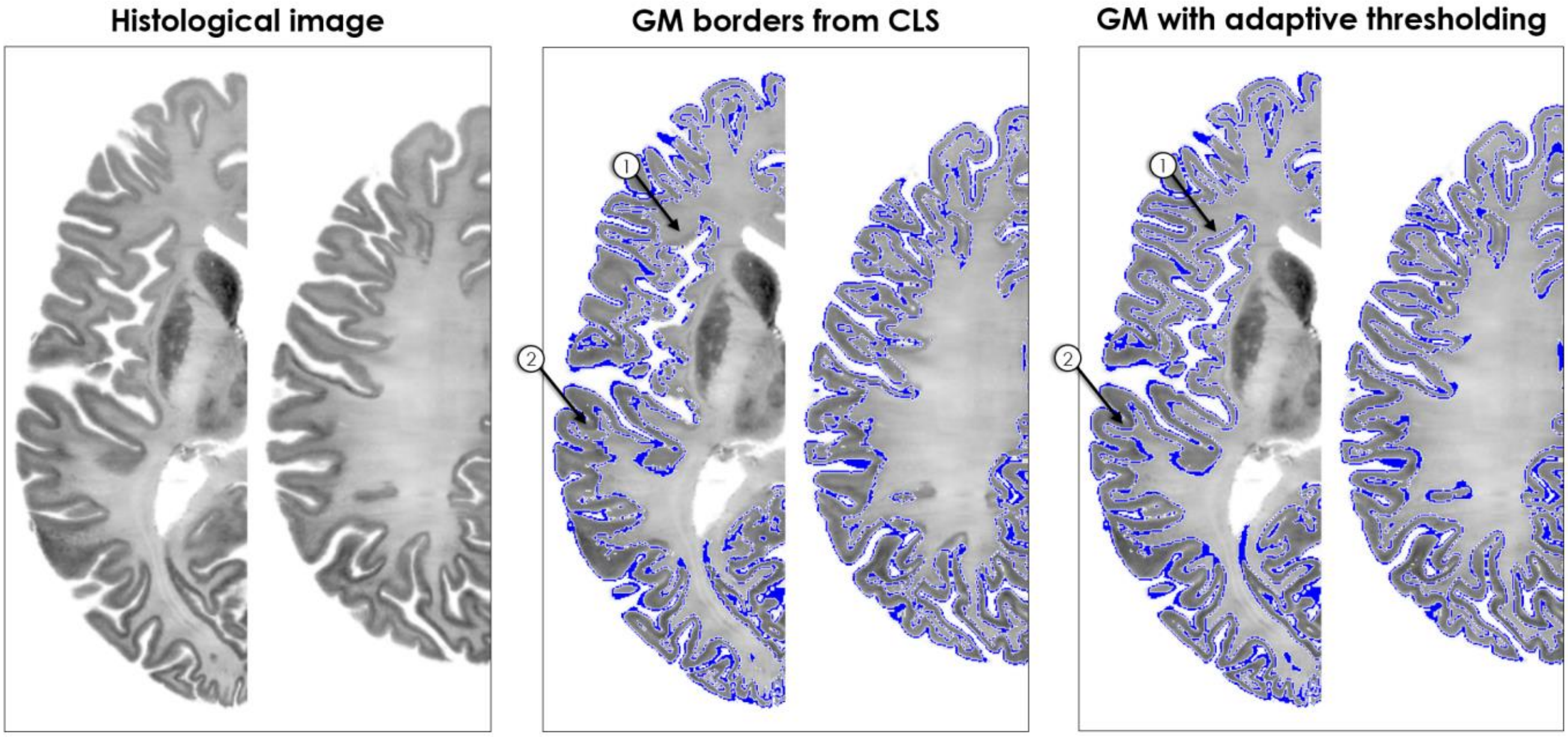

Supp. Fig. 1. Example illustration of the improvements obtained with the locally-adaptive thresholding approach implemented in this work for GM/WM differentiation. In this approach, the BigBrain classification volume ("cls") labels for GM and WM were used as a starting point. Then, for each GM/WM border voxel, an estimate of the local GM intensity was obtained by averaging the histological image intensities of voxels within a 4-mm radius that were labeled as GM in the cls; an analogous estimate was obtained for WM; the voxel was then assigned the label for which the estimated local average intensity was closest to that voxel's own intensity. For illustration, points 1 and 2 indicate regions where the GM had been under- and over-estimated in the cls, respectively; in 1 the GM is particularly lighter than in other regions, and thus closer in intensity to WM, whereas in 2 both GM and WM are particularly darker than in other regions. The adaptive approach deals with both cases more correctly.

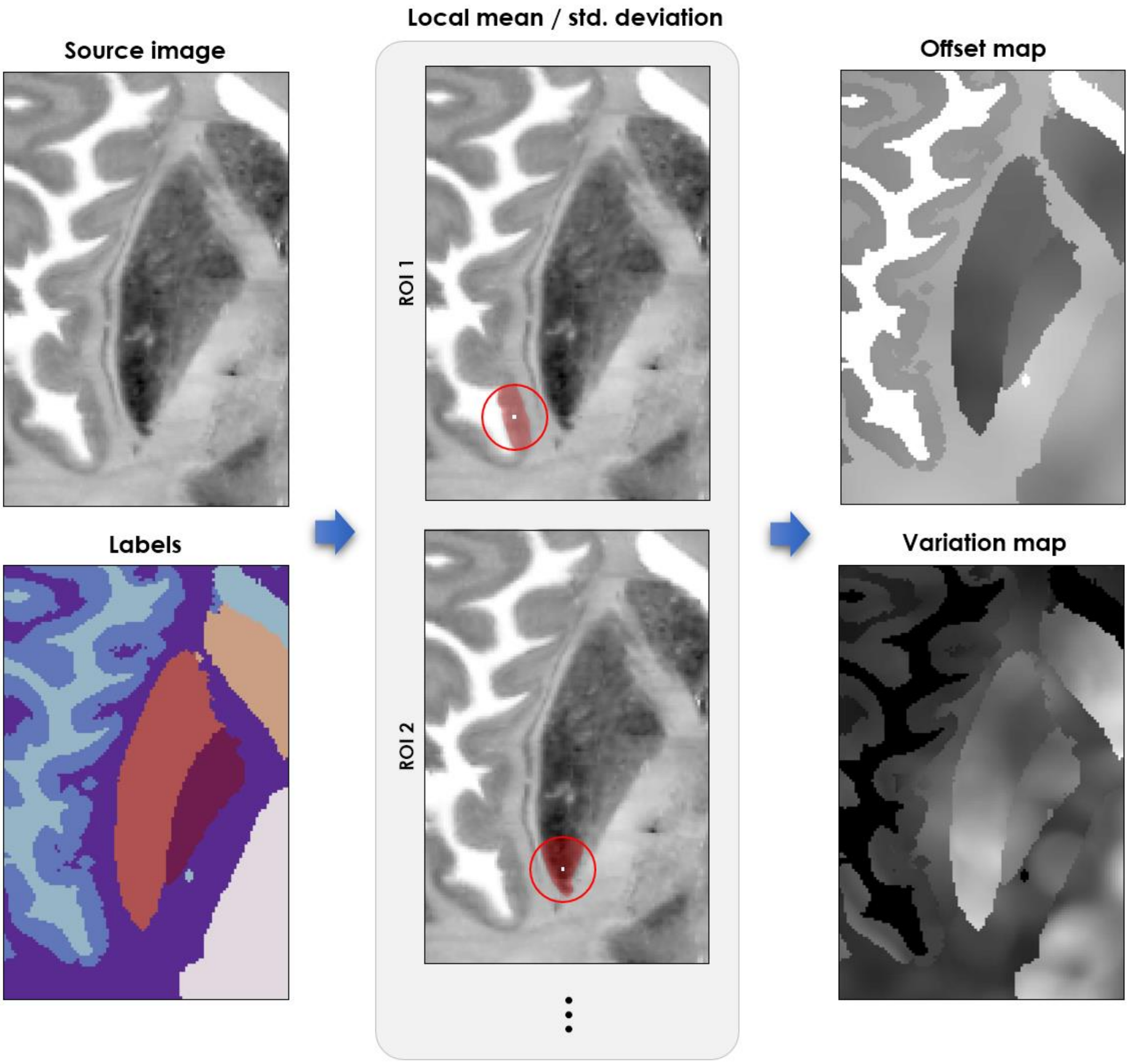

Supp. Fig. 2. Schematic illustration of the approach adopted in this work to obtain maps of ROI-specific local mean intensity (termed *offset*) and standard deviation (termed *variation*) from an arbitrary source image (in-vivo or histological), as described in detail in section 2.1.4. In this example, the BigBrain histological image is used as source. Two example ROIs are shown – GM and putamen. In each of the ROIs, the red ring represents the sliding spherical window, here centered on an example voxel (marked with a white dot). The region shaded red indicates the voxels included in the local average and local standard deviation estimates for that voxel. The resulting offset map essentially represents a smoothed version of the input, with the smoothing being confined within each ROI. The variation map reflects local heterogeneity in the intensities (resulting from “texture”, or other features, or noise) within each ROI.

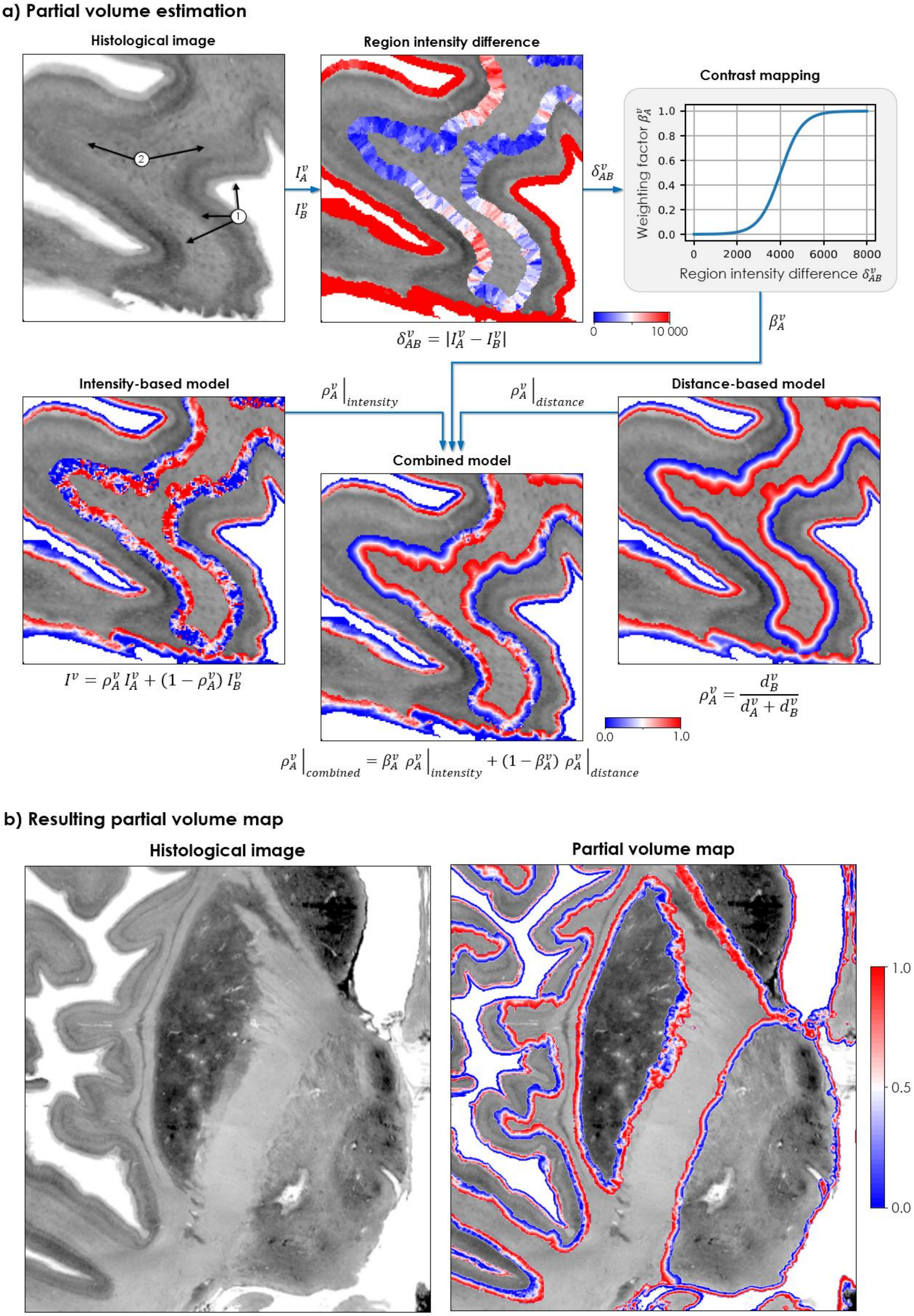
a) Partial volume estimation
Histological image
Region intensity difference
Contrast mapping
Weighting factor $\beta_A^v$
1.0
0.8
0.6
0.4
0.2
0.0
0
2000
4000
6000
8000
Region intensity difference $\delta_{AB}^v$
$I_A^v$
$I_B^v$
$\delta_{AB}^v$
0
10 000
$\delta_{AB}^v = |I_A^v - I_B^v|$
$\beta_A^v$
Intensity-based model
$\rho_A^v\big|_{intensity}$
$\rho_A^v\big|_{distance}$
Distance-based model
Combined model
$I^v = \rho_A^v I_A^v + (1-\rho_A^v) I_B^v$
$\rho_A^v = \frac{d_B^v}{d_A^v + d_B^v}$
0.0
1.0
$\rho_A^v\big|_{combined} = \beta_A^v \rho_A^v\big|_{intensity} + (1-\beta_A^v) \rho_A^v\big|_{distance}$
b) Resulting partial volume map
Histological image
Partial volume map
1.0
0.5
0.0

Supp. Fig. 3. a) Schematic illustration of the approach adopted in this work to estimate a partial volume map for the borders between the labeled regions, as described in detail in section 2.1.5. The approach included an intensity-based model and a distance-based model; the former, expected to be the most accurate, was meant to take precedence in borders where the neighboring regions have sufficiently different histological image intensities (e.g. borders indicated by point 1); the latter was meant to take over in borders where the intensities are too similar (e.g. those indicated by point 2), wherein the intensity-based partial volume estimates become too "noisy" and inaccurate. Accordingly, to combine the two models, a region intensity difference map was first estimated to quantify these differences along every border, and the two model contributions were then weighted at each voxel by a sigmoid function of the intensity difference. For the examples indicated by points 1 and 2, as illustrated, the former were thereby mostly reliant on the intensity-based model, whereas the latter were dominated by the distance-based model. b) Example region with the final estimated partial volume map, including well-defined borders between regions of distinct intensity (e.g. GM vs. background, putamen vs. WM) and less well defined borders (e.g. thalamus vs. WM, certain parts of GM vs. WM).

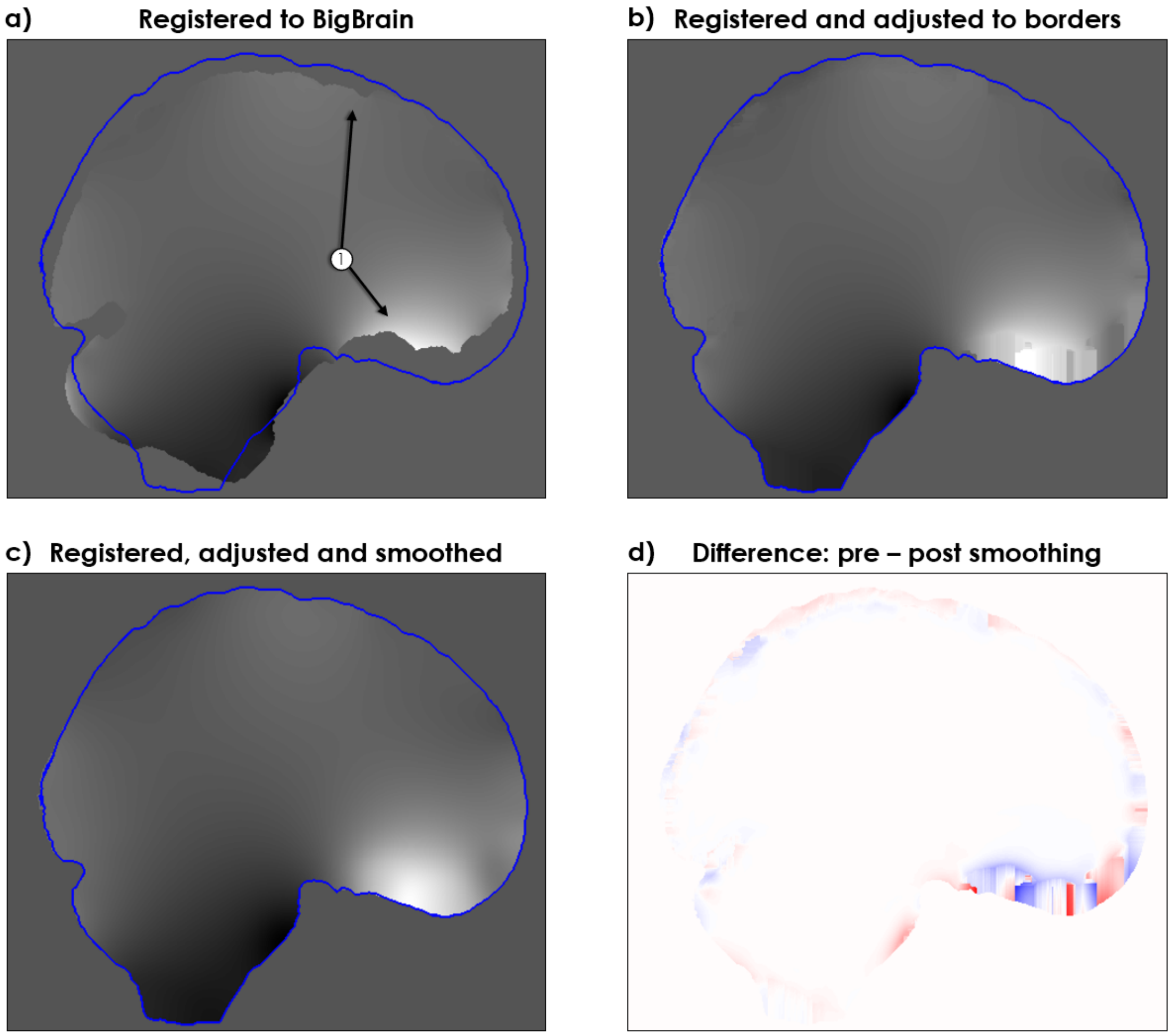

Supp. Fig. 4. Example illustration of the alternative mapping procedure that was employed for the coil sensitivity maps, background field map (which is shown in this example) and bias field maps. As described in 2.1.7, first, the maps were linearly registered to ICBM / BigBrain space (relying on magnitude images from the same acquisitions) at 400-µm resolution (a). Subsequently, where necessary, the maps had to be adjusted to adequately cover the BigBrain anatomy: regions that extended beyond the BigBrain borders were simply removed; regions that did not reach the BigBrain borders (as the examples indicated by point 1) were filled by expanding the existing map, whereby

each empty voxel bordering with filled voxels was assigned the same value as its closest filled neighbor, and the process was iteratively repeated until all gaps were covered (b). Finally, a wide spatial smoothing step (Gaussian, 8-mm full-width-at-half-maximum) was applied to remove small local discontinuities in the gap regions resulting from the expansion procedure (c); it could be confirmed by visual inspection that this smoothing step effectively cleared those discontinuities without altering the source regions (d).

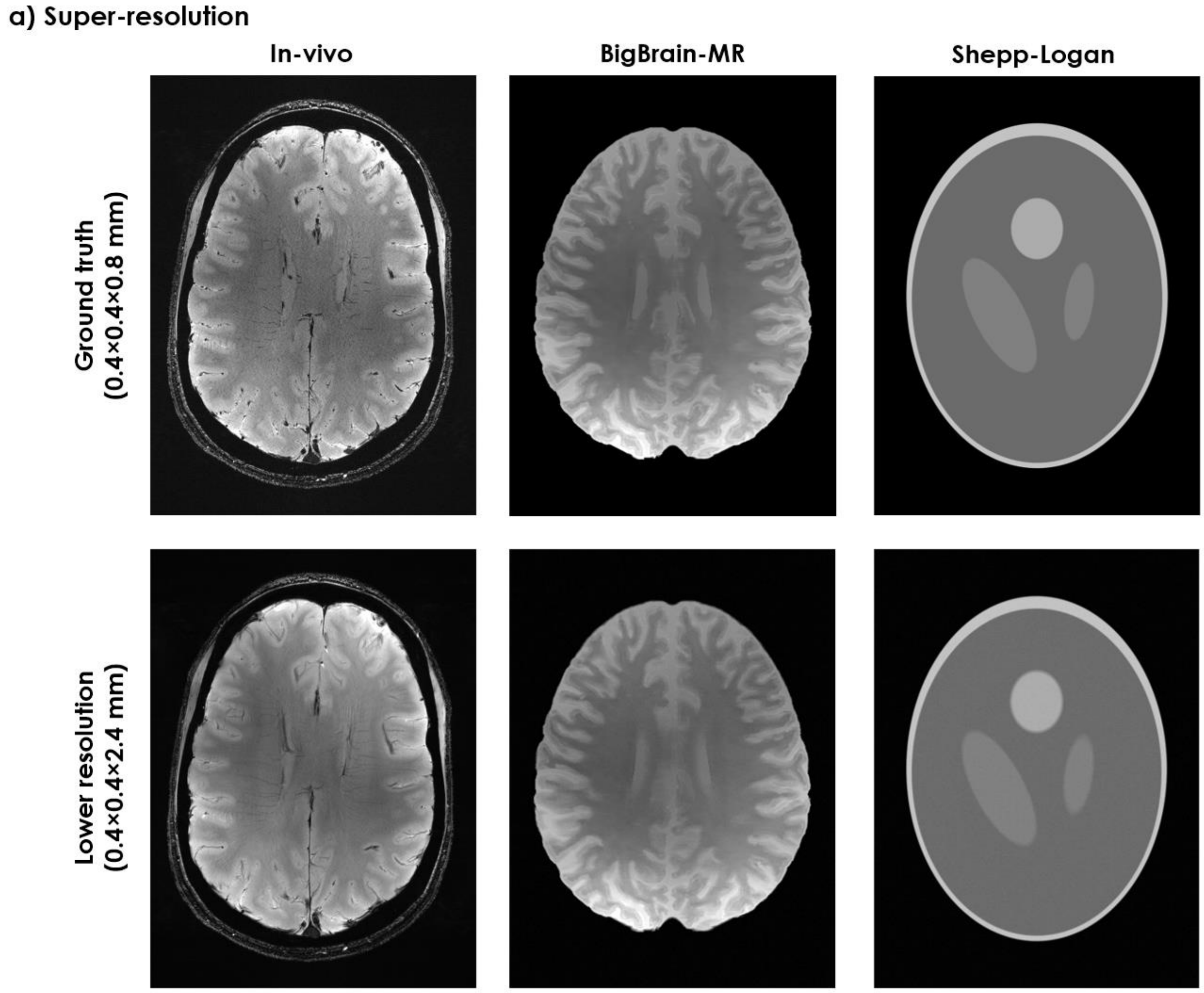

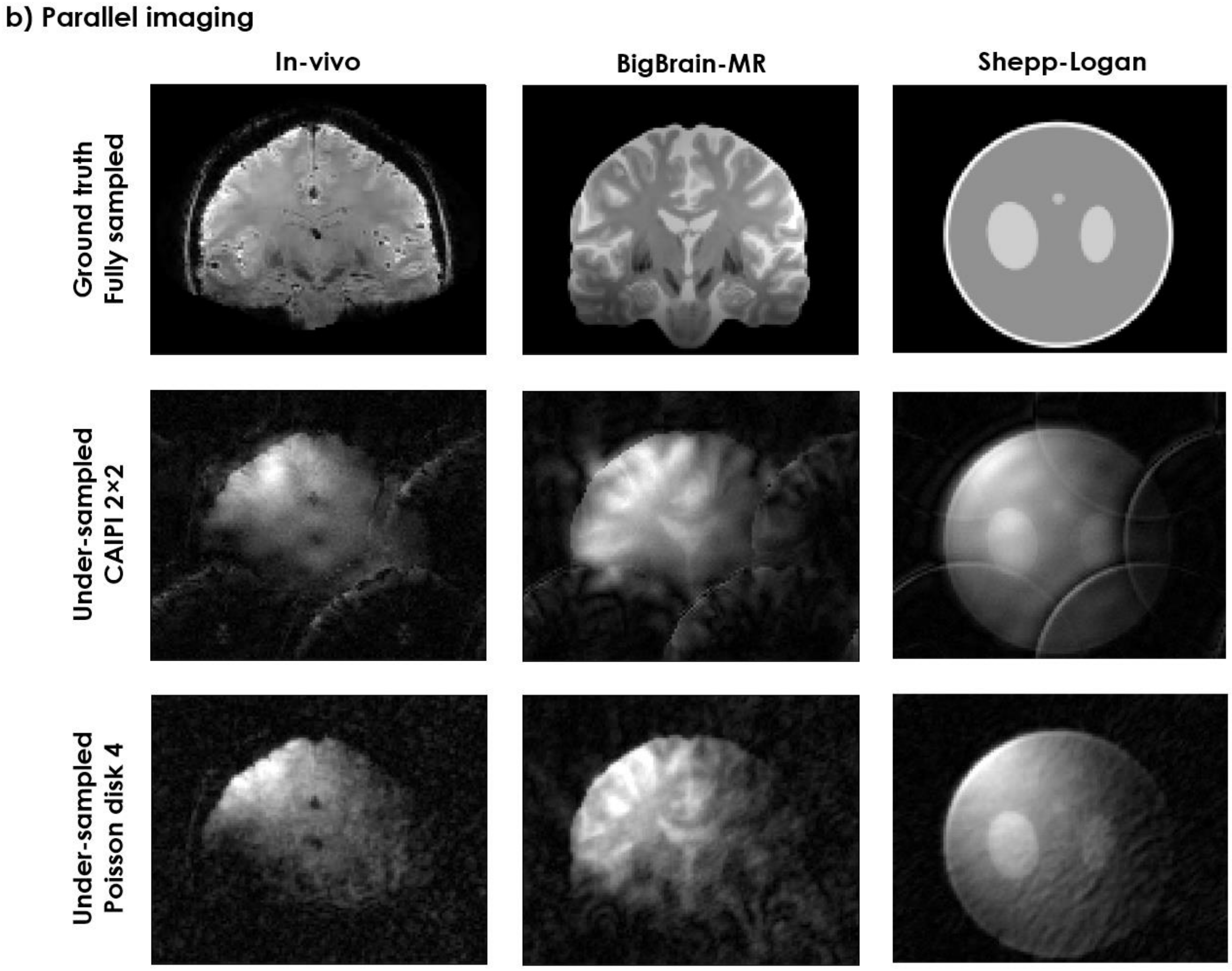

Supp. Fig. 5. Examples of the acquired and simulated images used for (a) the super-resolution imaging tests, and (b) the parallel imaging reconstruction tests. In (b), the k-space-undersampled images (2nd and 3rd row) are shown for one example channel for the RF receive array.

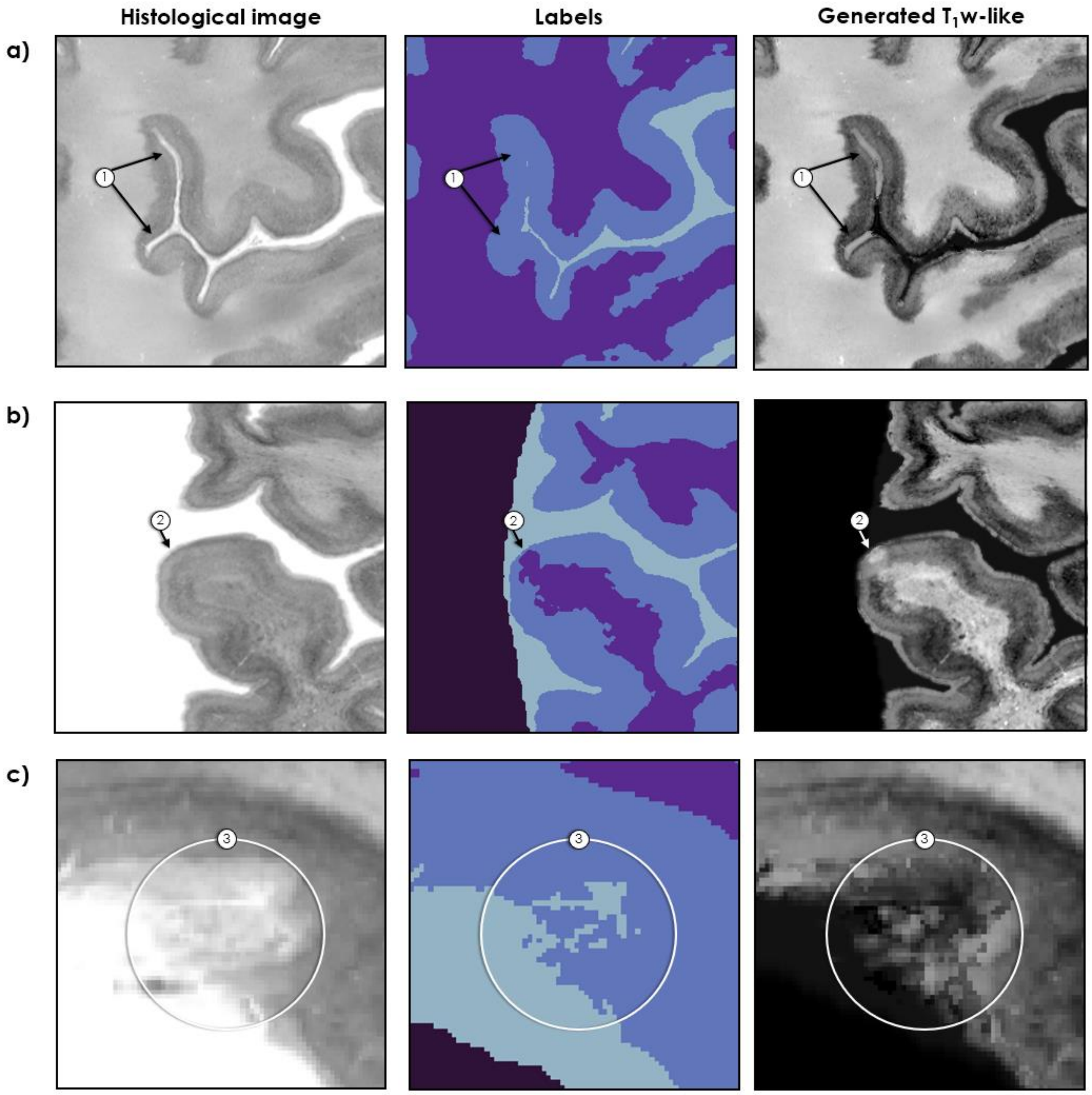

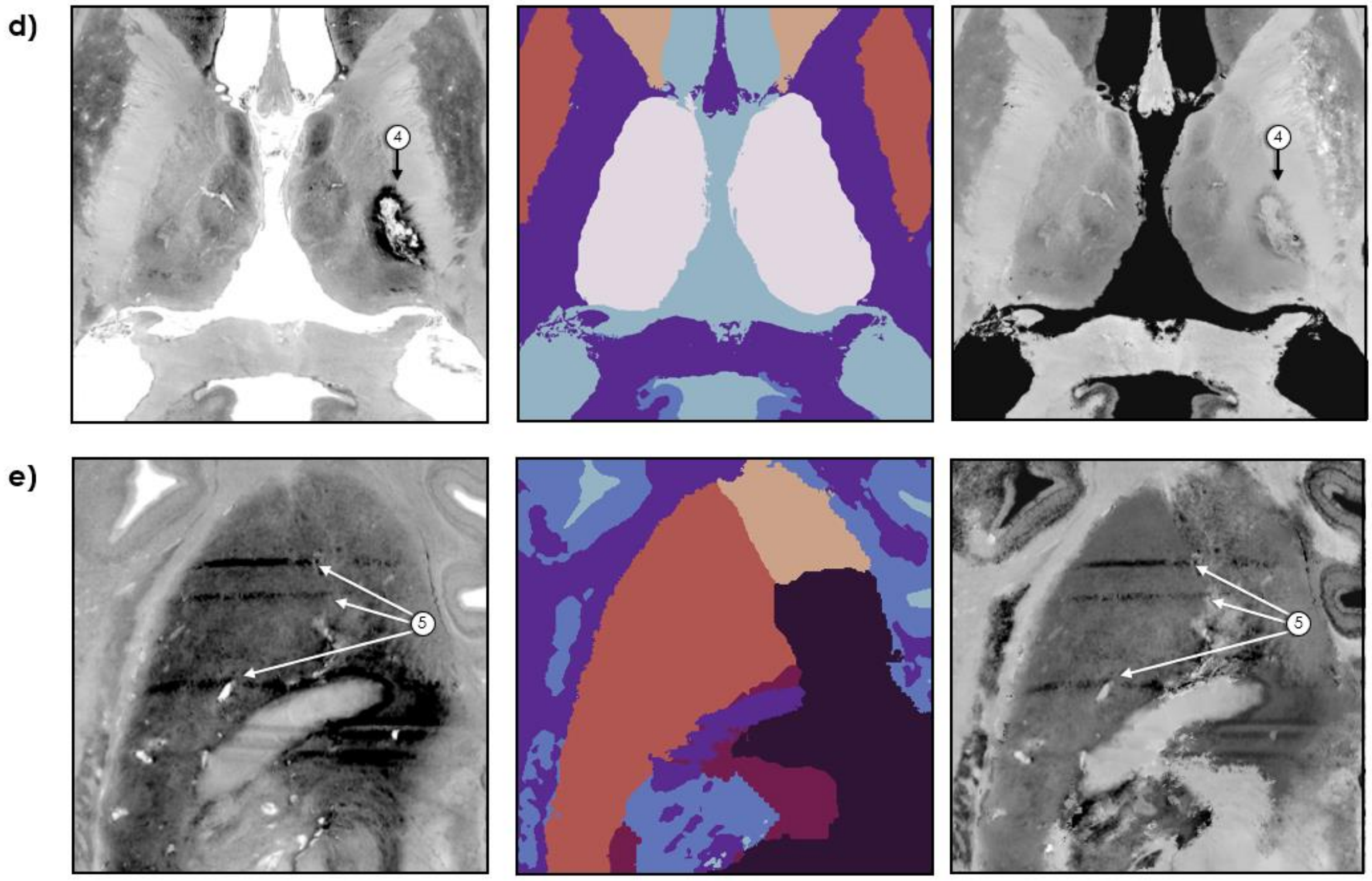

Supp. Fig. 6. Examples of local imperfections observed on the 100-µm BigBrain-MR images generated by the mapping framework, including flaws resulting from region labeling errors, often in difficult regions of the cortex such as very narrow sulci (a) and areas with unclear GM/WM boundary (b), artifacts introduced by the partial volume estimation, particularly in areas with complex/irregular/flawed borders (c), and also artifacts originated from the BigBrain histological preparation itself, such as damage to the thalamus (d) and staining artifacts (e).